\documentclass[12pt]{article}
\pdfoutput=1

\usepackage{rotating, verbatim}
\usepackage{epsfig}
\usepackage{latexsym}
\usepackage{graphicx}
\usepackage{psfrag}
\usepackage{amsmath,amssymb}
\usepackage{amsfonts}
\usepackage{graphicx}
\graphicspath{ {figures/} }
\usepackage{epstopdf}
\usepackage{color}
\usepackage{cite}
\usepackage{slashed}
\usepackage{hyperref}
\usepackage{datetime}
\usepackage[dvipsnames]{xcolor}

\newcommand{\be}{\begin{equation}}
\newcommand{\ee}{\end{equation}}
\newcommand\bea{\begin{eqnarray}}
\newcommand\eea{\end{eqnarray}}

\begin{document}

\thispagestyle{empty}

\begin{flushright}
DESY 15-246
\end{flushright}

\begin{center}

{\Large \bf Science with the space-based interferometer eLISA. \\
II: Gravitational waves from cosmological phase transitions}

\vspace{0.8cm}

{\large Chiara Caprini$^a$, Mark Hindmarsh$^{b,c}$,
Stephan Huber$^b$, \\Thomas Konstandin$^d$, Jonathan Kozaczuk$^e$, 
Germano Nardini$^f$, \\Jose Miguel No$^b$, Antoine Petiteau$^g$, 
Pedro Schwaller$^d$, \\G\'{e}raldine Servant$^{d,h}$, David J.
Weir$^i$\\} 
\vspace{0.5cm}
{\small $^a$ IPhT, CEA Saclay and CNRS UMR3681, 91191 Gif-sur-Yvette, France\\
$^b$ Department of Physics and Astronomy, University of Sussex, BN1 9QH, Brighton, UK\\
$^c$ Department of Physics and Helsinki Institute of Physics,
PL 64, 
FI-00014 University of Helsinki,
Finland\\
$^d$ DESY, Notkestrasse 85, D-22607 Hamburg, Germany\\
$^e$ TRIUMF, 4004 Wesbrook Mall, Vancouver, BC V6T 2A3, Canada \\
$^f$ ITP, AEC, University of Bern,
Sidlerstrasse 5, CH-3012, Bern, Switzerland\\
$^g$ APC, Universit\'e Paris Diderot, Observatoire de Paris, Sorbonne Paris Cit\'e,
10 rue Alice Domon et L\'eonie Duquet, 75205 Paris Cedex 13, France \\
$^h$ Institute of Theoretical Physics, Univ. Hamburg, D-22761 Hamburg, Germany\\
$^i$ Institute of Mathematics and Natural Sciences, University of Stavanger, 4036 Stavanger, Norway }
\\[1cm]

\end{center}

{
\abstract{
We investigate the potential for the eLISA space-based interferometer to 
detect the stochastic gravitational wave background produced by strong 
first-order cosmological phase transitions. We discuss the resulting 
contributions from bubble collisions, magnetohydrodynamic turbulence, and sound waves to the 
stochastic background, and 
estimate the total corresponding signal predicted in gravitational waves. The projected sensitivity of eLISA to cosmological phase 
transitions is computed in a model-independent way for various detector designs and configurations. By applying these results to several specific models, we 
demonstrate that eLISA is able to probe many well-motivated 
scenarios beyond the Standard Model of particle physics predicting strong 
first-order cosmological phase transitions in the early Universe.}
}

\tableofcontents

\newpage

\section{\sc Introduction}

Phase transitions (PTs) are ubiquitous in nature. Boiling liquid into gas, the 
emergence of superconductivity, superfluidity, and permanent magnetization
in ferromagnetic materials are some well-known examples of this phenomenon. 
Intriguingly, PTs can also be \emph{cosmological}: regions of the 
universe can abruptly  transition from one ground state to 
another. In the language of quantum field theory, this typically corresponds to 
the appearance or change of a scalar field operator's vacuum expectation value 
(VEV). 

A \emph{first-order} cosmological PT can occur when two local 
minima of the free energy co-exist for some range of temperatures. If this is 
the case, the relevant scalar field can quantum mechanically `tunnel'  or 
thermally fluctuate into the new phase. These quantum or thermal processes proceed via the 
nucleation of bubbles in a sea of metastable phase. The bubbles will then expand 
and eventually collide with each other. This sequence of events can give rise to 
a significant stochastic background of gravitational waves (GWs), as we describe below, and provides an 
attractive target for the next space-based GW
observer, the eLISA interferometer \cite{elisaweb}.

First-order cosmological PTs are predicted in many scenarios 
beyond the Standard Model of particle physics. They can be tied to the 
production of the observed baryon asymmetry, to the nature of dark matter, or 
simply be a byproduct of an extended scalar sector. Consequently, eLISA can be a 
vital tool in exploring possibilities for new physics, complementing existing 
observational efforts at colliders, precision and cosmic frontier experiments. 

The purpose of this paper is to quantitatively assess the extent to which eLISA 
can test realistic scenarios predicting strong first-order PTs in 
the early Universe. 
Currently, a design study is in progress within the European Space
Agency to define the most scientifically promising configuration of eLISA. The characteristics of the
configuration that remain open or undetermined are the low-frequency noise
level, soon to be tested by the LISA Pathfinder \cite{pathfinderweb},
the number of laser links (four or six, corresponding to two or three
interferometer arms), the length of the interferometer arms (1, 2, or 5
million km), and the duration of the mission (2 or 5 years). 

As part of the design study, the scientific potential of each possible
eLISA configuration is currently being analyzed and scrutinized.    
This paper is the second in a series that evaluates the impact of the aforementioned four key design specifications on the scientific performance of eLISA. The first considered the primary scientific target of eLISA: the GW signal from massive black hole binaries \cite{Klein:2015hvg}. 
In this paper, we specifically address
eLISA's potential to detect a stochastic GW background arising from a first-order PT. As we will see, our conclusions in this regard depend quite sensitively on the 
experimental configuration.
For a more detailed discussion on the status of the eLISA mission and its possible designs, we refer the reader to the first paper of this series \cite{Klein:2015hvg}. 

The remainder of this study is structured as follows. We 
provide an overview of GW generation at a first-order 
cosmological PT in Section~\ref{sec:signal}. 
Model-independent projections of the eLISA sensitivity to such signals are 
presented in Section~\ref{sec:sensitivity}. We then discuss several specific 
examples of models predicting strong GW signatures through a 
first-order PT in Section~\ref{sec:models}, either associated with electroweak symmetry
breaking (Section~\ref{sec:EWPT_models}) or otherwise (Section~\ref{sec:beyond}). 
Section~\ref{sec:conc} comprises a summary and our conclusions. Some additional details of our analysis
are provided in Appendix~\ref{app:alphainfty}.

\subsection{Definitions and Notation} \label{sec:def}
Before proceeding, we comment briefly on our notation, which generally coincides with that found in 
e.g. Ref.~\cite{Grojean:2006bp}, and identify the quantities important for computing the GW signal
from cosmological PTs. 

In what follows, $T_*$ denotes the temperature of 
the thermal bath at the time $t_*$ when GWs are produced. For typical transitions without significant reheating, 
this is approximately equivalent to the nucleation temperature,  $T_*\approx T_n$. For the remainder of this section we assume that this is the case, deferring the treatment of scenarios with large reheating effects to Section~\ref{sec:case3}.
The bubble nucleation rate is \cite{Turner:1992tz}
\be
\Gamma (t) =A(t)e^{-S(t)} \, , 
\ee
where $S$ is the 
Euclidean action of a critical bubble\footnote{In principle, there are many solutions to the Euclidean 
equations of motion that can contribute to the action. In practice, only the solution with the 
lowest action is relevant. For vacuum transitions, this is the
$O(4)$-symmetric solution with Euclidean action $S_4$, while at finite
temperature the 
$O(3)$-symmetric bounce is the relevant solution, with Euclidean action $S_3/T$. Then, at a given time, $S(t)\approx \operatorname{min}\{S_4, S_3/T\}$.}. $T_n$ is then defined as the temperature at which $\Gamma$ becomes large enough to nucleate a bubble per horizon volume with probability of order 1.

In terms of $\Gamma(t)$, the (approximate) inverse time duration of the PT 
is defined as
\be
\beta \equiv - \left.\frac{dS}{dt}\right|_{t=t_*} \simeq   \frac{\dot{\Gamma}}{{\Gamma}}
 \, .
\ee
A key parameter controlling the GW signal is
the fraction $\beta/H_*$, where $H_*$ is the Hubble parameter at $T_*$. 
The smaller $\beta/H_*$ is, the stronger the phase transition and consequently 
the GW signal. Provided $T_*\approx T_n$, this ratio can be expressed as
\be
\label{eq:boH_thermal}
\frac{\beta}{H_*}=T_* \left. \frac{dS}{dT}\right|_{T_*} \, .
\ee

A second key
parameter, $\alpha$, is the ratio of the vacuum energy density released in the transition to that of the radiation bath~\footnote{
Notice that in parts of the literature, $\alpha$ is defined in terms of the 
latent heat of the PT and not in terms of the 
vacuum energy. In the limit of strong PTs and large supercooling these 
two definitions coincide.}. For transitions without significant reheating,
\be
\label{eq:alpha_thermal}
\alpha=\frac{\rho_{\rm vac}}{\rho_{\rm rad}^*} \, ,  
\ee
where $\rho_{\rm rad}^*=g_*\pi^2 T_*^4/30$, 
and $g_*$ is the number of relativistic degrees of freedom in the plasma at $T_*$. 

The characteristic frequency of the signal at the time of emission is set roughly by $H_*$ (see Section \ref{sec:contributions}).  
For PTs taking place at the electroweak epoch, the observed frequency today, which is redshifted by a factor of $(T_0/T_*) $, is typically in the milliHertz range.  eLISA, which is sensitive to frequencies in this range, may therefore provide a window to electroweak scale physics.

Other important  parameters  are \cite{Binetruy:2012ze}
\be
   \kappa_v=\frac{\rho_{v}}{\rho_{\rm vac}}\,,  \ \   \ \ \kappa_\phi=\frac{\rho_\phi}{\rho_{\rm vac}}\,,
\ee
the fraction of vacuum energy that gets converted into bulk motion of the fluid and into gradient energy of the Higgs-like field, respectively. 
Finally, $v_w$ denotes the bubble wall velocity in the rest frame of the fluid far away from the bubble. 

\section{\sc Prediction of the Gravitational Wave Signal}\label{sec:signal}

As mentioned above, a first-order PT proceeds by the nucleation 
and expansion of bubbles of a new phase. Isolated spherical bubbles 
 do not source GWs. It
is therefore processes arising during the collision of bubbles that
are relevant for GW physics (we ignore possible
instabilities~\cite{Megevand:2013yua,Megevand:2014yua}).

Depending on the details of the model, the bubble walls can move with
velocities up to the speed of light. If the dynamics of the PT involve only a single field with no relevant couplings to
the thermal bath, then the bubble walls move relativistically in what
is known as a vacuum transition.

Thermal PTs, in which the scalar field is coupled to a
plasma of light fields, typically involve slower wall velocities due
to the effective friction term coupling the field to the
plasma~\cite{Espinosa:2010hh}. It is, however, still possible to get
luminal wall velocities~\cite{Bodeker:2009qy}. For thermal PTs, 
if velocities are subsonic $v_w<1/\sqrt{3}$, a shock forms in the plasma in front of the bubble wall; if velocities are supersonic $v_w>1/\sqrt{3}$, a rarefaction wave forms behind the bubble wall. 

The spectrum of the stochastic GW background arising from cosmological PTs depends on 
various sources. These are outlined in Section~\ref{sec:contributions}. Which sources are most relevant in a given PT 
scenario depends sensitively on the dynamics of bubble expansion. We discuss the various possibilities in Section~\ref{sec:scenarios}. 

\subsection{Contributions to the Gravitational Wave Spectrum}\label{sec:contributions}

To varying degrees, three processes are involved
in the production of GWs at a first-order PT:
\begin{itemize}
\item Collisions of bubble walls and (where relevant) shocks
  in the plasma. These can be treated by a technique now generally
  referred to as the `envelope 
approximation'~\cite{Kosowsky:1991ua,Kosowsky:1992rz,astro-ph/9211004,
astro-ph/9310044,Caprini:2007xq,Huber:2008hg}. As described below, this approximation can be used to compute the contribution to the GW spectrum from the 
scalar field, $\phi$, itself.
\item Sound waves in the plasma after the bubbles have
  collided but before expansion has dissipated the kinetic energy in
  the plasma~\cite{Hindmarsh:2013xza,Giblin:2013kea,Giblin:2014qia,Hindmarsh:2015qta}.
\item Magnetohydrodynamic (MHD) turbulence in the plasma forming after the 
bubbles have collided~\cite{Caprini:2006jb,Kahniashvili:2008pe,arXiv:0909.0622,Kisslinger:2015hua,Kahniashvili:2009mf,Kahniashvili:2008pf}.
\end{itemize}
\noindent
These three processes generically coexist, and the corresponding contributions to the stochastic GW background
should linearly combine, at least approximately, so that
\begin{equation}\label{eq:total2}
h^2\Omega_{\rm GW} \simeq h^2\Omega_{ \phi} + h^2\Omega_{\rm sw} + 
h^2\Omega_{\rm turb}\,.
\end{equation}
Let us briefly review each contribution in more detail.

\subsubsection{Scalar Field Contribution}
\label{sec:scalar}

The GW contribution from the scalar field involved in the PT can be treated using the envelope approximation \cite{Kosowsky:1991ua,Kosowsky:1992rz,astro-ph/9211004,astro-ph/9310044}. 
In this approximation, a fraction $\kappa$ of the latent heat of the PT is deposited in a thin shell close to the PT front.
The energy in each shell is then assumed to quickly disperse after colliding with another shell such that the energy is primarily stored in the envelope of 
uncollided shells \footnote{The envelope approximation we adopt here neglects the fact that the scalar field can perform oscillations as it settles into the true vacuum after wall collisions, as demonstrated e.g. in \cite{Child:2012qg}.}. Numerical simulations utilizing the envelope approximation suggest that the GW contribution to the spectrum is given by~\cite{Huber:2008hg}
\bea
h^2\Omega_{\rm env}(f) &=& 1.67 \times 10^{-5}   \, \left( \frac{H_*}{\beta} \right)^2 \left( \frac{\kappa \alpha}{1+\alpha} \right)^2  
 \left( \frac{100}{g_*} \right)^{\frac{1}{3}} \left(\frac{0.11\,v_w^3}{0.42+v_w^2}\right) \, S_{\rm env}(f) \, ,
\label{eq:Omenv}
\eea
where $S_{\rm env}(f)$ parametrises the spectral shape of the GW radiation. A fit to simulation data~\cite{Huber:2008hg} yields
\be\label{Senv}
S_{\rm env}( f ) = \frac{3.8 \,\,(f/f_{\rm env})^{2.8}}{1 + 2.8 \, (f/f_{\rm env})^{3.8}}\, ,
\ee
with the slopes of the spectrum in the limit of small and large frequencies given respectively by $S_{\rm env}\propto f^q$ with $q = 2.8$ and $S_{\rm env}\propto f^{-p}$ with $p=1$. Causality implies that at low frequency the spectral index is $q=3$ \cite{Caprini:2009fx}. This has to be the case at least for frequencies smaller than the inverse Hubble horizon at GW production, Eq.~\eqref{Hstar}. However, $q=2.8$ provides a better fit to the simulated result close to the peak of the spectrum and we adopt this spectral index in the following. 

The peak frequency of the contribution to the spectrum from bubble collisions, $f_{\rm env}$, is determined by the characteristic time-scale of the PT, i.e.~its duration $1/\beta$ \cite{Caprini:2009fx,arXiv:0909.0622}. From simulations, the peak frequency (at $t_*$) is approximately given by~\cite{Huber:2008hg}
\be
\frac{f_*}{\beta} = \left(\frac{0.62}{1.8-0.1v_w+v_w^2} \right) \, .
\label{eq:fstar}
\ee
This value is then red-shifted to yield the peak frequency today \cite{astro-ph/9310044}, 
\be
f_{\rm env} = 16.5 \times 10^{-3} \, {\rm mHz} \, \left(\frac{f_*}{\beta}\right) \, \left(\frac{\beta}{H_*} \right) 
\left(\frac{T_*}{100\,{\rm GeV}}\right) \left( \frac{g_*}{100}
\right)^{\frac{1}{6}}\,.
\label{eq:envPeak}
\ee
In going from Eq.~\eqref{eq:fstar} to Eq.~\eqref{eq:envPeak} we use the value of the inverse Hubble time at GW production, redshifted to today,
\be
h_* = 16.5 \times 10^{-3} \, {\rm mHz} 
\left(\frac{T_*}{100\,{\rm GeV}}\right) \left( \frac{g_*}{100} 
\right)^{\frac{1}{6}}
\label{Hstar}
\ee
along with the assumption that the Universe transitioned directly to a radiation-dominated phase after the PT and has expanded adiabatically ever since.

The envelope approximation can be readily applied to the GW contribution 
arising from the scalar field itself,
\be
\label{eq:OmegaPhi}
h^2 \Omega_{\phi}(f) = \left. h^2 \Omega_{\rm env}(f) 
\right|_{\kappa = \kappa_\phi} \, ,
\ee
where $\kappa_\phi$ denotes the fraction of latent heat transformed 
into the kinetic energy of the scalar field. Its size depends on the details of the bubble expansion, as we discuss in Section \ref{sec:scenarios}.


\subsubsection{Sound Waves} 

Percolation produces bulk motion in the fluid in the form of sound waves. 
Acoustic production of GWs is an area of active research, and a definitive model covering all 
relevant $v_w$ and $\alpha$ is not yet available \cite{Hindmarsh:2013xza,Hindmarsh:2015qta}.  
For generic values of $v_w$ (meaning values more than 
about 10\% away from the sound speed or the speed of light), the numerical results of \cite{Hindmarsh:2015qta} 
are fitted reasonably by
\begin{equation}
\begin{aligned}
h^2\Omega_{\rm sw}(f) = 2.65 \times 10^{-6} \, \left( \frac{H_*}{\beta} \right) \left( \frac{\kappa_v \alpha}{1+\alpha} \right)^2  
 \left( \frac{100}{g_*} \right)^{\frac{1}{3}} v_w \, S_{\rm sw}(f) \, ,
\label{eq:OmGsound} 
\end{aligned}
\end{equation}
where the efficiency $\kappa_v$ denotes the fraction of latent heat that is transformed into bulk motion of the fluid, and depends on the expansion mode of the bubble (see Section \ref{sec:scenarios}).

The numerical simulations performed in Ref.~\cite{Hindmarsh:2015qta} indicate that 
the contribution from acoustic production can be modelled by a broken power law, with the causal slope $q=3$ for values of the 
frequency below the peak frequency, and a power law $-p$ above the peak, with $p \gtrsim 3$. It can be shown that the signal-to-noise ratios for GW detection are rather insensitive to the precise value of $p$ if it is greater than 3. For the purposes of this analysis we take $p=4$. 
We adopt the following spectral shape $S_{\rm sw}( f )$ in Eq.~\eqref{eq:OmGsound}:
\be\label{Ssw}
S_{\rm sw}( f ) = (f/f_{\rm sw})^{3}\,\left(\frac{7}{4 + 3\,(f/f_{\rm sw})^{2}}  \right)^{7/2}\, .
\ee

The peak frequency $f_{\rm sw}$ is not yet well understood. 
The overall scale is set by the average bubble separation $R_* = (8\pi)^\frac13 v_w/\beta$, but the peak position is found numerically to be less than $R_*$. 
A conservative estimate that complies with the above spectral shape is  $f_{\rm sw}=(2/\sqrt{3})(\beta/v_w)$, which, after redshifting, becomes 
\be
f_{\rm sw} = 1.9 \times 10^{-2} \, {\rm mHz} \, \frac{1}{v_w} \, \left(\frac{\beta}{H_*} \right) 
\left(\frac{T_*}{100\,{\rm GeV}}\right) \left( \frac{g_*}{100} 
\right)^{\frac{1}{6}}\,.
\label{eq:SWPeak}
\ee

The parametric dependence of the GW spectrum in Eq.~\eqref{eq:OmGsound} differs by a factor $\beta/H_*$ with respect to the envelope result in Eq.~\eqref{eq:Omenv}. The enhancement of the spectral amplitude by a factor $\beta/H_*$ for long-lasting sources w.r.t to short lasting ones has been predicted on the basis of analytical arguments in Ref.~\cite{arXiv:0909.0622}. Simulations show that the sound waves typically remain active as a source of GW much longer than the collisions 
of the bubble walls \cite{Hindmarsh:2015qta}. We therefore believe that the $\beta/H_*$ factor is tied to the duration of the source, and is robust. The same amplification is observed in the case of MHD turbulence which is also a long-lasting source (it takes several Hubble times to dissipate \cite{arXiv:0909.0622}).

We emphasize that the simulations used to arrive at Eqs.~(\ref{eq:OmGsound})--(\ref{eq:SWPeak}) were restricted to values of $\alpha\lesssim 0.1$ and the maximum root mean square fluid velocity $\sqrt{\kappa\alpha_v}$ was about $0.05$. The extent to which the results of these simulations can be extrapolated to larger $\alpha$ remains to be investigated. 
In particular, we expect that the development of weak shocks at $t_{\rm sh} \sim (v_w/\sqrt{\kappa\alpha_v})\beta^{-1}$ (see e.g.\  \cite{Pen:2015qta}) will eventually convert the acoustic signal to a turbulent one, described in the next section.
We urge the reader to keep this in mind when interpreting our results below.

\subsubsection{MHD Turbulence} 

Percolation can also induce turbulence in the plasma, and in particular MHD turbulence since the plasma is fully ionized. 
Considering Kolmogorov-type turbulence as proposed in Ref.~\cite{Kosowsky:2001xp}, 
its contribution to the GW spectrum in Eq.~\eqref{eq:total2} can be modelled as\footnote{Note that MHD turbulence after a primordial PT can also be helical, as pointed out e.g. in Ref.~\cite{Kahniashvili:2008pe}. Here we neglect the GW signal from a possible helical component, which in the context of eLISA will be the subject of a subsequent study.} \cite{arXiv:0909.0622,Binetruy:2012ze} 
\bea
\label{eq:turb1}
h^2\Omega_{\rm turb}(f) = 
3.35 \times 10^{-4} \, \left( \frac{H_*}{\beta} \right)
\left(\frac{\kappa_{\rm turb}\,\alpha}{1+\alpha}\right)^{\frac{3}{2}}\,
 \left( \frac{100}{g_*}\right)^{1/3}\, v_w \, S_{\rm turb} (f) \, ,
\eea
where $\kappa_{\rm turb}$ denotes the fraction of latent heat that is transformed 
into MHD turbulence (note the different dependence on this parameter w.r.t to the sound wave and scalar field cases). Similarly to the case of sound waves, one recognizes the amplification by a factor $\beta/H_*$ which is typical of sources that last longer than the average duration $1/\beta$ of the PT. 
The spectral shape has been found analytically and is given by \cite{arXiv:0909.0622,Binetruy:2012ze} 
\be
S_{\rm turb} (f) = \frac{(f/f_{\rm turb})^3}
{\left[ 1 + (f/f_{\rm turb}) \right]^{\frac{11}{3}} 
\left(1 + 8 \pi f/h_* \right)} \,. 
\label{Sturb}
\ee
The explicit dependence on the Hubble rate $h_*$ (c.f.~Eq.~\eqref{Hstar}) is also a consequence of the fact the turbulence acts as a source of GW for several Hubble times. The causal slope $q=3$ is recovered for frequencies smaller than $h_*$, but it changes to $q=2$ for intermediate (sub-Hubble) values of the frequency $h_*<f<f_{\rm turb}$, again due to the long duration of the GW source \cite{arXiv:0909.0622}. At large frequency $f\gg f_{\rm turb}$ the slope is determined by the Kolmogorov turbulence model, $p=5/3$. 

Similarly to the sound wave case, the peak frequency is connected to the inverse characteristic length-scale of the source, the bubble size  $R_*$ towards the end of the PT. Analytic arguments show that this is due to the particular time de-correlation properties of the turbulent source \cite{arXiv:0909.0622}. One has $f_{\rm turb}\simeq (3.5/2) (\beta/v_w)$, which becomes, after redshifting,
\be\label{eq:TurbPeak}
f_{\rm turb} = 2.7 \times 10^{-2} \, {\rm mHz} \, \frac{1}{v_w}\, 
\left(\frac{\beta}{H_*} \right) \left(\frac{T_*}{100\,{\rm GeV}}\right) \left( 
\frac{g_*}{100} \right)^{\frac{1}{6}} \, . 
\ee
Note that, contrary to the envelope and sound wave cases, the spectral shape of the GW spectrum in Eq.~\eqref{Sturb} has not been tuned such that $f_{\rm turb}$ exactly corresponds to the maximum of $S_{\rm turb} (f)$. Here $f_{\rm turb} $ is inherited from the physical arguments underlying the analytical evaluation carried out in Ref.~\cite{arXiv:0909.0622}.  

\vskip 3 mm

\subsection{Dynamics of the Phase Transition: Three Cases} \label{sec:scenarios}
The relative importance of each contribution discussed above w.r.t.~GW generation 
depends strongly on the features of the PT and its dynamics. 
The bubble wall velocity plays a key role in this respect. 
If the wall velocity is small, the thermal bath can efficiently absorb the
energy available in the PT, and the GW spectrum
is thus suppressed. If the wall velocity is instead relativistic, 
a large portion of the energy budget can go into bulk motion 
or even in the kinetic energy of the wall itself. 
Moreover, in the case of relativistic bubble wall velocities, 
qualitatively different scenarios can arise, depending on whether the bubble
wall reaches a terminal velocity or not. In the latter (`runaway')
case, a further important distinction is whether or not plasma effects play an
important role in bubble expansion. Below, we consider these
different scenarios in turn~\footnote{We do not analyze the case
of non-relativistic wall velocity since it tends to yield GW spectra
that are not detectable in the forthcoming future.}. We stress that, when 
analyzing the GW spectrum predicted in a particular particle physics
model, it is important to understand which of the following scenarios apply.


\subsubsection{Case 1: Non-runaway Bubbles \label{sec:case1}}
Bubbles expanding in a plasma can reach a relativistic terminal velocity. In this case, 
the energy in the scalar field is negligible (it only scales with the surface of the bubble and not with the volume)
and the most relevant
contributions to the signal are expected to arise from the bulk motion
of the fluid. This can be in the form of sound waves and/or
MHD turbulence. Combining these contributions, we approximate the total
spectrum as
\begin{equation}\label{eq:total1}
h^2\Omega_{\rm GW} \simeq h^2\Omega_{\rm sw} + h^2\Omega_{\rm turb} \,.
\end{equation}
%

As shown in the previous sections, these expressions involve $\kappa_v$, the efficiency factor for conversion of 
the latent heat into bulk motion \cite{astro-ph/9310044,Espinosa:2010hh}. In 
the limits of small and large $v_w$, it is approximately given by
\begin{equation}
\label{eq:kappav}
\kappa_v \simeq 
\left\{\begin{array}{c c}
\alpha \left(0.73+0.083\sqrt{\alpha}+\alpha\right)^{-1} & v_w \sim 1\\
v_w^{6/5}6.9\, \alpha \left(1.36-0.037 \sqrt{\alpha}+\alpha \right)^{-1},& v_w 
\lesssim 0.1
\end{array}
\right.~
\end{equation}
Full expressions for $\kappa_v$ are given in Ref.~\cite{Espinosa:2010hh}, which 
we utilize below (note that $\kappa_v$ is called $\kappa$ in 
Ref.~\cite{Espinosa:2010hh}). 

The GW spectra also depend on $v_w$, which is model-dependent. We choose $v_w=0.95$ 
for concreteness, since scenarios with nearly luminal wall velocities are more 
promising from the standpoint of observable gravitational radiation.

In the GW contribution from MHD turbulence, Eq.~\eqref{eq:turb1}, we
take 
\be\label{epsilon}
\kappa_{\rm turb} = \epsilon\, \kappa_v \, ,
\ee
 with $\epsilon$ 
representing the fraction of
bulk motion which is turbulent.  Recent simulations suggest that only at most 
$5-10\%$ of the bulk motion from the bubble walls is converted into
vorticity (cf.~e.g.~Table II in \cite{Hindmarsh:2015qta}). 
However, these simulations lasted for less than one eddy turn-over time so one would not expect significant turbulence to have developed. 
The onset of turbulence is expected after shocks develop at $t_{\rm sh} \sim (v_w/\sqrt{\kappa\alpha_v})\beta^{-1}$, which is less than a Hubble time for stronger transitions. More work is needed to understand how turbulence develops from the acoustic waves, and to allow for the uncertainty 
in what follows we conservatively set $\epsilon=0.05$. This strongly suppresses the role of turbulence as far as the detection of GW from the PT is concerned, thereby underestimating the signal in the case that weak shocks develop within one Hubble time.
As we will see, turbulence can only slightly improve the signal-to-noise ratio in extreme cases for which the PT is very slow, i.e. $\beta\simeq H_*$. 
A more accurate balance between acoustic and turbulent gravitational wave production
remains to be investigated, as does the possible contribution from the magnetic field.

\subsubsection{Case 2: Runaway Bubbles in a Plasma} \label{sec:case2}

If a model predicts a first-order PT already at the
mean-field level, it is possible for the bubble wall to accelerate
without bound and hence run away~\cite{Bodeker:2009qy}, with
$v_w\rightarrow c$. Although the existence of a runaway configuration
does not guarantee that it will be realized~\cite{Konstandin:2010dm,
  Kozaczuk:2015owa}, it is generally difficult to prevent strong
transitions from reaching the runaway regime, if it exists.

Bubbles can run away even if expanding in a thermal plasma. In this case, the 
energy density stored in the Higgs-like field profile cannot be neglected, since it 
is known to dominate as $\alpha\rightarrow \infty$ (Case 3 below). The total 
contribution to the GW signal can be approximated in this case 
by
\begin{equation}
h^2\Omega_{\rm GW} \simeq h^2\Omega_{ \phi} + h^2\Omega_{\rm sw} + 
h^2\Omega_{\rm turb}\,,
\end{equation}
where we have reintroduced $\Omega_{\rm \phi}$, the part sourced by gradients in 
the scalar field. This contribution is well-modeled by the envelope approximation (see Section~\ref{sec:scalar}).

The following picture emerges in this case~\cite{Espinosa:2010hh}. As $\alpha$ is
increased, the wall velocity quickly becomes relativistic. We denote by
$\alpha_{\infty}$ the minimum value of $\alpha$ such that bubbles
run away (i.e. no longer reach a terminal velocity in the plasma frame).  
For $\alpha>\alpha_{\infty}$, the fraction of the total phase transition energy budget deposited into the fluid saturates.
Beyond this value, the fluid profile no longer changes with increasing $\alpha$ and the
surplus energy goes into accelerating the bubble wall. This surplus energy is parameterized by the fraction
\be
\kappa_{\phi}\equiv \frac{\alpha-\alpha_{\infty}}{\alpha} \ge 0\, .
\ee
Only the fraction $\alpha_{\infty}/\alpha$ of the total energy budget is then transformed into bulk
motion and thermal energy according to Eq.~(\ref{eq:kappav}):
\begin{equation}
\begin{aligned}
&\kappa_v \equiv \frac{\alpha_{\infty}}{\alpha} \kappa_{\infty}\, , \label{eq:kapv} 
\\
&\kappa_{\rm therm} \equiv 
\left(1-\kappa_{\infty}\right)\frac{\alpha_{\infty}}{\alpha}\,,\\
&\kappa_{\infty} \equiv 
\frac{\alpha_{\infty}}{0.73+0.083\sqrt{\alpha_{\infty}}+\alpha_{\infty}}\,.
\end{aligned}
\end{equation} 

The parameter $\alpha_{\infty}$ is model-dependent. Denoting the
tunneling field\footnote{Here, $\phi$ should be thought of as a vector in field space for scenarios in which more than one field changes its VEV during the transition.} as $\phi$ with corresponding vacuum expectation value $\phi_*$ inside the bubble immediately after tunneling, one can express
$\alpha_{\infty}$ as~\cite{Espinosa:2010hh}
\be
\alpha_{\infty} \simeq \frac{30}{24 \pi^2}\frac{\sum_a c_a \Delta m_a^2(\phi_*)}{g_* T_*^2}\, ,
\ee
for the typical case with $T_*\approx T_n$. In the above expression, the sum runs over all particles $a$ that are light in the
initial phase and heavy in the final phase, $\Delta m_a^2(\phi_*)$ is
the difference of their (field-dependent) squared masses in the two phases,
$g_*$ again corresponds to the effective number of relativistic degrees of freedom in
the initial phase, and $c_a$ is equal to $N_a$ for bosons and
$\frac{1}{2} N_a$ for fermions, with $N_a$ the number of the degrees
of freedom for the species $a$. For electroweak PTs in models
with Standard Model-like particle content ($g_*=106.75$, $a\in
\left\{W^{\pm}, Z, t \right\}$ and $N_W=6$, $N_Z=3$, $N_t=12$)~\footnote{The Higgs and Goldstone
    contributions are typically numerically negligible.}, this
  parameter is approximately given by~\footnote{Note that the expression
  above differs from that in Ref.~\cite{Espinosa:2010hh}; we believe
  this discrepancy is due to an algebraic error in the latter.}
\begin{equation}\label{eq:alinf}
\alpha_{\infty}\simeq 4.9\times 10^{-3} \left(\frac{\phi_*}{T_*}\right)^2\, .
\end{equation}
Since most scenarios beyond the Standard Model do not feature new
relativistic degrees of freedom or new particles with couplings to the
Higgs comparable to those of the $SU(2)_L$ gauge bosons or top quark,
Eq.~\eqref{eq:alinf} is typically a reliable estimate for electroweak
PTs, although one should verify this explicitly with the
exact expressions found in Ref.~\cite{Espinosa:2010hh}.

By definition, reasonably strong first-order PTs in a radiation-dominated epoch feature
$\phi_*/T_*\gtrsim 1$. On the other hand, heuristically, models with
$\phi_*/T_* \gtrsim 10$ are more likely to fall into the
large-$\alpha$ case (described in the next section) for which $\alpha_{\infty}$
becomes irrelevant. We therefore expect $\alpha_{\infty} \sim 0.005$ --
0.5 for scenarios belonging to Case 2. 

\subsubsection{Case 3: Runaway Bubbles in Vacuum} \label{sec:case3}

Finally, some models predict PTs that occur in a vacuum-dominated epoch. This situation arises in models with a significant amount of supercooling. In this case, plasma effects are negligible, and the bubble wall will accelerate indefinitely, with $v_w$ quickly approaching the speed of light, as in Case 2 above. 

The temperature $T_*$ in this situation is given approximately by the reheat temperature after percolation, $T_*\approx T_{\rm reh}$. So far, in most of our expressions we have assumed $T_*\approx T_n\approx T_{\rm reh}$, as is appropriate for transitions in a radiation-dominated epoch and without significant reheating. However, in the vacuum case, one instead generally expects $T_n \ll T_{\rm reh} \approx T_*$, since $T_{\rm reh}$ is governed by the vacuum energy released during the transition. The definitions of $\alpha$ and $\beta/H_*$ should be adjusted accordingly. In particular, for vacuum transitions, Equations~(\ref{eq:boH_thermal}) and~(\ref{eq:alpha_thermal}) should be replaced by
\be
\label{eq:boH_vacuum}
\frac{\beta}{H_*}=\frac{H(T_n)}{H_*}T_n \left. \frac{dS}{dT}\right|_{T_n}, \qquad \alpha = \frac{\rho_{\rm vac}}{\rho_{\rm rad}(T_n)}.
\ee
Notice that for fast reheating one obtains $H(T_n) \simeq H_*$ even though $T_n \ll T_*$. This is 
because energy conservation ensures that the vacuum energy that dominates $H(T_n)$ is transformed without loss into the radiation energy that dominates $H_*$.

As $T_n\rightarrow 0$, $\alpha\rightarrow \infty$ and the $\alpha$ dependence drops out of the predicted GW signal in this scenario (c.f.~Eq.~\eqref{eq:Omenv}). 
Also, in this limit, only the Higgs field contribution is significant, from which it follows that
\begin{equation}\label{eq:total3}
h^2\Omega_{\rm GW} \simeq h^2\Omega_{ \phi} \, ,
\end{equation}
where $h^2\Omega_{ \phi}$ is given in Eq.~\eqref{eq:OmegaPhi}, and Eq.~(\ref{eq:Omenv}) with 
$\kappa_{\phi}=1$, $v_w=1$. There is  no significant plasma contribution in this 
case,  by definition. Note that, if the reheating of the Standard Model sector is slow, there will be a period of matter domination immediately following the transition, which would change the redshift, and hence Eq.~(\ref{eq:envPeak}). We will not consider this particular case further.

\section{ \sc eLISA Sensitivity} \label{sec:sensitivity}

\subsection{Detection Threshold}
\label{sec:decthr}

In this analysis we consider four representative configurations for eLISA, which we name C1-C4 and which are listed in Table~\ref{tab:configurations}. The corresponding eLISA sensitivity curves can be found in Ref.~\cite{Klein:2015hvg} for the target GW source, massive black hole binaries. On the other hand, here we are interested in a stochastic background of GWs, which is statistically homogeneous and isotopic. The purpose of this section is to briefly explain how one can obtain sensitivity curves which correctly represent the prospects for detecting  a stochastic GW background with eLISA for a given configuration (more details will be presented in an upcoming study~\cite{AntoineP}).

\begin{table}[th!]
\begin{center}
\begin{tabular}{|l|c|c|c|c|}
\hline
Name & C1 & C2 & C3 & C4 \\ \hline
Full name & N2A5M5L6 & N2A1M5L6 & N2A2M5L4 &  N1A1M2L4 \\ \hline
\# links & 6 & 6 & 4 & 4 \\ \hline
Arm length [km] & 5M & 1M & 2M & 1M \\ \hline
Duration [years] & 5 & 5 & 5 & 2 \\ \hline
Noise level  & N2 & N2 & N2 & N1 \\ \hline
\end{tabular}
\caption{\label{tab:configurations} Properties of the representative
  eLISA configurations chosen for this study.  The corresponding
  sensitivity curves are shown in Figure~\ref{fig:sensitivities}. More
  details on these configurations and their sensitivity curves can be
  found in Ref.~\cite{Klein:2015hvg} and Ref.~\cite{AntoineP}
  respectively. }
\end{center}
\end{table}%

\begin{figure}
\begin{center}
\includegraphics[width=0.7\textwidth]{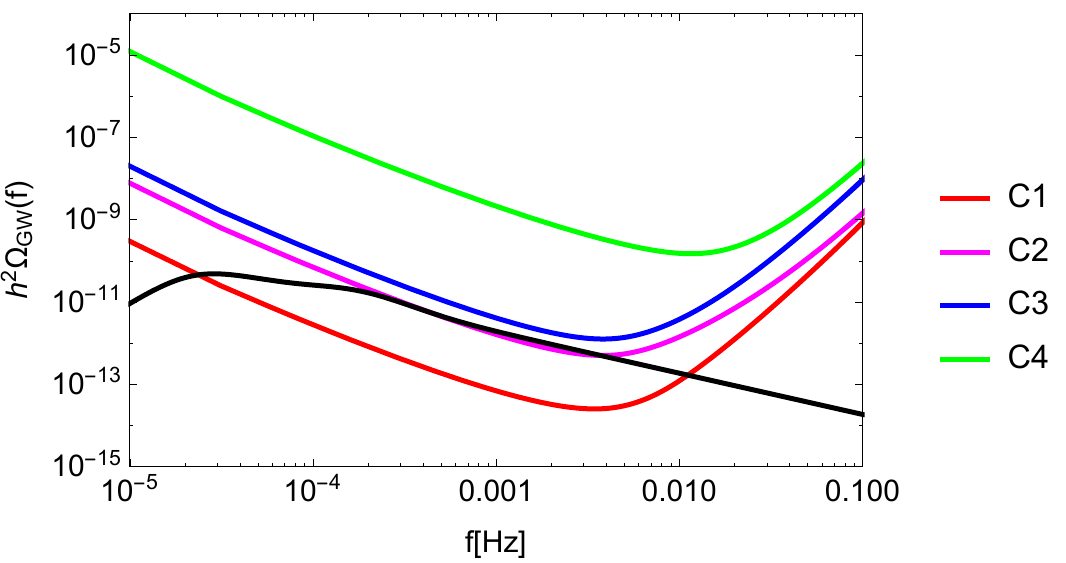}
\caption{\small \label{fig:sensitivities} Sensitivity curves of the C1-C4 configurations given in Table~\ref{tab:configurations} compared with a typical GW signal. We have chosen the signal predicted in the Higgs portal scenario described in Section \ref{sec:Hportal}, with benchmark values $T_*=59.6$, $\alpha=0.17$, $\beta/H_*=12.54$, $\phi_*/T_*=4.07$ (see Table \ref{table:scalars}). } 
\end{center}
\end{figure}

For the C1-C4 configurations, the resulting eLISA sensitivity to a stochastic GW background is shown in Figure~\ref{fig:sensitivities}. 
The most promising clearly appears to be C1, which corresponds to the old LISA configuration: it has 6 links, 5
million km arm length, a duration of 5 years and noise level
corresponding to that expected to be found by the LISA Pathfinder (labeled as N2 and henceforth called ``LISA Pathfinder expected''). The least sensitive is
C4, with 4 links, 1 million km arm length, a duration of 2 years
and noise level corresponding to 10 times larger than expected (N1, also dubbed ``LISA Pathfinder
required''). 
For the intermediate configurations, we have fixed the
duration to five years and the noise level to LISA Pathfinder expected, 
since these two characteristics are likely achievable. An open
question, which we would like to answer with this analysis, is whether
it is more efficient to add a pair of laser links or to increase the arm length
for the purpose of probing the occurrence of first-order PTs in the early Universe. {\it The outcome, as we will see, is
that adding a pair of laser links leads to a larger gain in sensitivity
 than increasing the arm length from 1 to 2 million
km.}

To assess the detectability of the GW signal, we consider the signal-to-noise 
ratio \cite{Thrane:2013oya},
\begin{equation}
{\rm SNR}=\sqrt{{\cal T} \int_{f_{\rm min}}^{f_{\rm max}} df\, \left[\frac{h^2 
\Omega_{\rm GW}(f)}{h^2 \Omega_{\rm Sens}(f)}\right]^2},
\label{SNR}
\end{equation}
where $h^2 \Omega_{\rm Sens}(f)$ denotes the sensitivity of a given eLISA 
configuration and ${\cal T}$ is the duration of the mission in years \cite{AntoineP}. Whenever ${\rm SNR}$ is larger than a threshold 
${\rm SNR_{thr}}$, the signal $h^2 \Omega_{\rm GW}(f)$ can be detected. 
Quantifying ${\rm SNR_{thr}}$ is not an easy task. We briefly describe how this can be done here, referring the interested reader to Ref.~\cite{AntoineP} for more details.

Applying a Bayesian method, Refs.~\cite{Adams:2010vc,Adams:2013qma} found that 
the old LISA configuration over one year can detect a white-noise 
stochastic background at the level of $h^2\Omega_{\rm GW}^{\rm L6} =1\times 10^{-13}$. 
This sensitivity can be achieved by exploiting the fact that, with three interferometer arms (i.e.~three pairs of laser links), it is possible to 
form two (virtually) noise-independent detectors, in which the noise is 
uncorrelated whereas the GW signal is correlated. This technique is safe and 
robust, although it remains to be tested with realistic noise levels. On the other 
hand, this technique cannot be applied to the two-arm configurations, and the 
level of detectable GW background is degraded. With the same Bayesian method, and 
assuming good prior knowledge of the noise, Ref.~\cite{Adams:2013qma} finds that 
with a four-link but otherwise LISA-like configuration over one year 
one can detect a white-noise stochastic background at the level of 
$h^2\Omega_{\rm GW}^{\rm L4} =3.5 \times 10^{-13}$. 

These levels of sensitivity $h^2\Omega_{\rm GW}^{\rm L6,L4}$ do account for the presence of the confusion noise from galactic binaries: Refs.~\cite{Adams:2010vc,Adams:2013qma} show that this latter can be estimated, and therefore extracted, together with the cosmological signal using the Bayesian method. This is why we do not include it in the sensitivity curves of the C1-C4 configurations shown in Fig.~\ref{fig:sensitivities}. 

For the present detection analysis, we use the above
results and convert them into corresponding values of ${\rm SNR_{thr}}$. We compare the
$h^2\Omega_{\rm GW}^{\rm L6}$ and $h^2\Omega_{\rm GW}^{\rm L4}$
detection levels with the power law sensitivity curve for each six-link
(respectively, four-link) configuration. The power law sensitivity curve is a
concept developed in \cite{Thrane:2013oya} with the aim of accounting for the improvement in the usual sensitivity curves of a GW detector that comes from the broadband nature of a stochastic
signal. The curve is given by the envelope of power laws $\Omega_\beta
(f/f_{\rm ref})^\beta$ that can be detected with ${\rm SNR}=1$,
varying $\beta$. For each eLISA configuration, we compute the power
law sensitivity curve, and the ${\rm SNR}$ corresponding to 
the detection levels $h^2\Omega_{\rm GW}^{\rm L6}$ and $h^2\Omega_{\rm
  GW}^{\rm L4}$. To be conservative, for the four-link configurations we
increase the detectability level to $h^2\Omega_{\rm GW}^{\rm
  L4}=10^{-12}$. This yields ${\rm SNR}=10$ for all six-link
configurations and ${\rm SNR}=50$ for all four-link configurations. We
then interpret these values as ${\rm SNR_{thr}}$. In other words, we classify a
given GW background $h^2 \Omega_{\rm GW}(f)$ as detectable by a six-link
configuration (four-link) if, once inserted into Eq.~\eqref{SNR},
it returns ${\rm SNR}>10$ (${\rm
  SNR}>50$) \cite{AntoineP}. Since we choose ${\rm SNR_{thr}}$ based on \cite{Adams:2010vc,Adams:2013qma}, the confusion noise from galactic binaries is accounted for in our analysis.

\subsection{Examples of Gravitational Wave Spectra}  
\label{sec:ex_spectra}
  
We have seen in Section \ref{sec:contributions} that the GW spectrum from a first-order PT is in general given by the sum of three contributions: that of the scalar field gradients, sound waves and MHD turbulence. The relative importance of each contribution depends on the details of the PT dynamics, as discussed in Section \ref{sec:scenarios}. Here we provide some examples of spectra to show the interplay among the aforementioned contributions. This helps to clarify the model-independent sensitivity contours derived in the next section. 

\begin{figure}
\begin{center}
\includegraphics[width=0.4\textwidth]{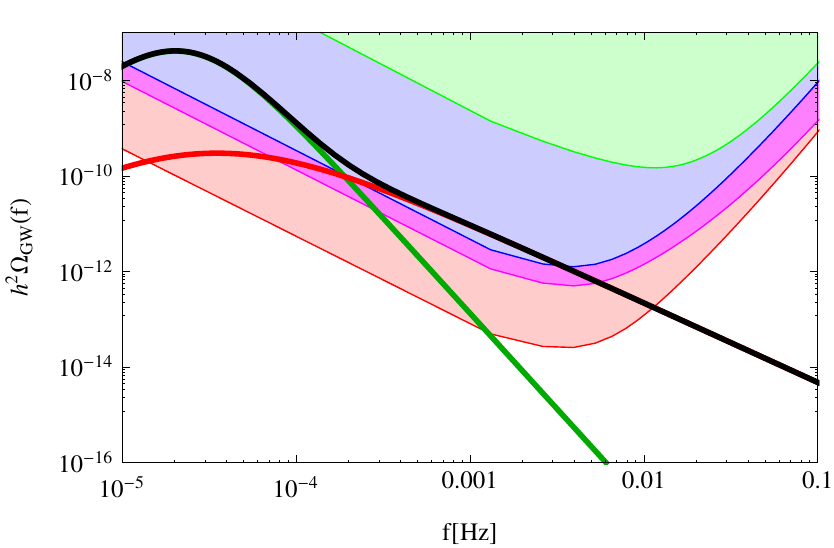}
\includegraphics[width=0.4\textwidth]{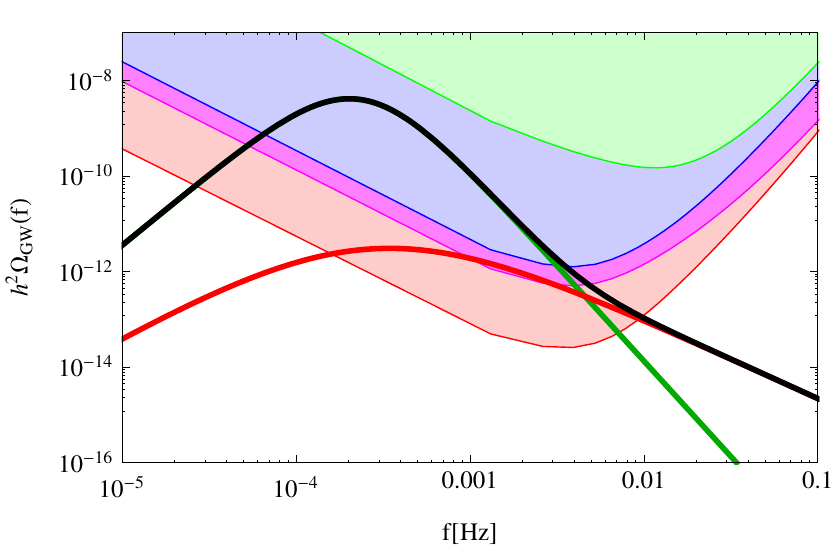}
\includegraphics[width=0.4\textwidth]{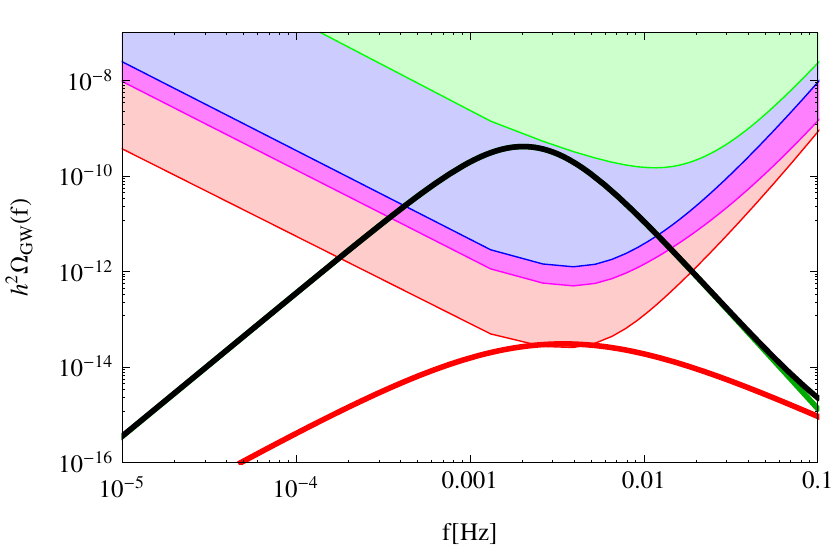}
\includegraphics[width=0.4\textwidth]{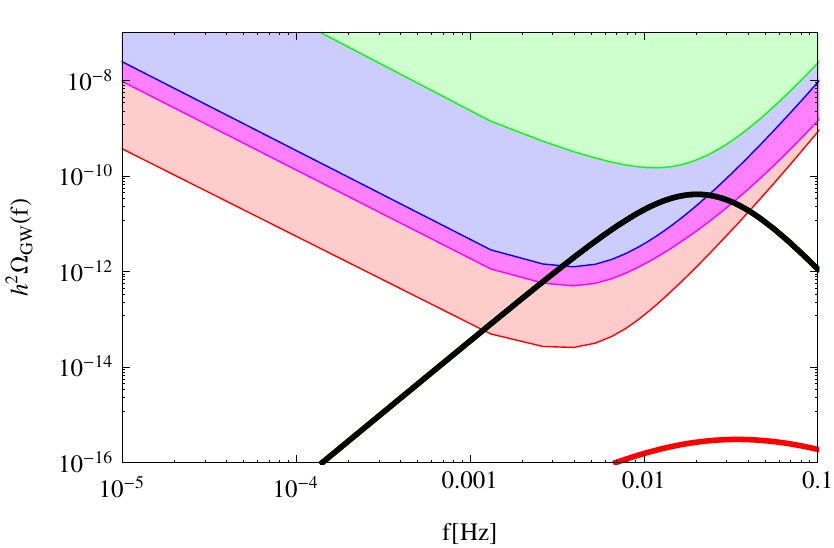}
\end{center}
\caption{\small \label{fig:spec_case1} Example of GW spectra in Case 1, for fixed $T_*=100$ GeV, $\alpha=0.5$, $v_w=0.95$, and varying $\beta/H_*$: from left to right, $\beta/H_*=1$ and $\beta/H_*=10$ (top), $\beta/H_*=100$ and $\beta/H_*=1000$ (bottom). The black line denotes the total GW spectrum, the green line the contribution from sound waves, the red line the contribution from MHD turbulence. The shaded areas represent the regions detectable by the C1 (red), C2 (magenta), C3 (blue) and C4 (green) configurations. }
\end{figure}

In Figure~\ref{fig:spec_case1} we show examples of GW spectra that can arise if the PT proceeds through non-runaway bubbles (Case 1 above), for fixed $T_*$, $\alpha$ and $v_w$, and varying $\beta/H_*$. The total spectrum is given in Eq.~\eqref{eq:total1} by the sum of the signal generated by sound waves and by MHD turbulence. Since we set $\epsilon=0.05$ the signal from sound waves is dominant (c.f.~Eq.~\eqref{epsilon}). Turbulence can play a role at high frequencies, because the signal from sound waves decays faster (i.e.~with $p=4$ as opposed to $p=5/3$, c.f.~Eqs.~\eqref{Ssw} and \eqref{Sturb}) and the peak positions are not that different (c.f.~Eqs.~\eqref{eq:SWPeak} and~\eqref{eq:TurbPeak}). Increasing $\beta/H_*$ at fixed $T_*$ and $v_w$ causes the peak position to shift towards larger frequencies, and the overall amplitude to decrease. Correspondingly, the contribution from turbulence becomes less and less important because of the suppression due to the factor $8\pi f/h_*$ in Eq.~\eqref{Sturb}. 

\begin{figure}
\begin{center}
\includegraphics[width=0.4\textwidth]{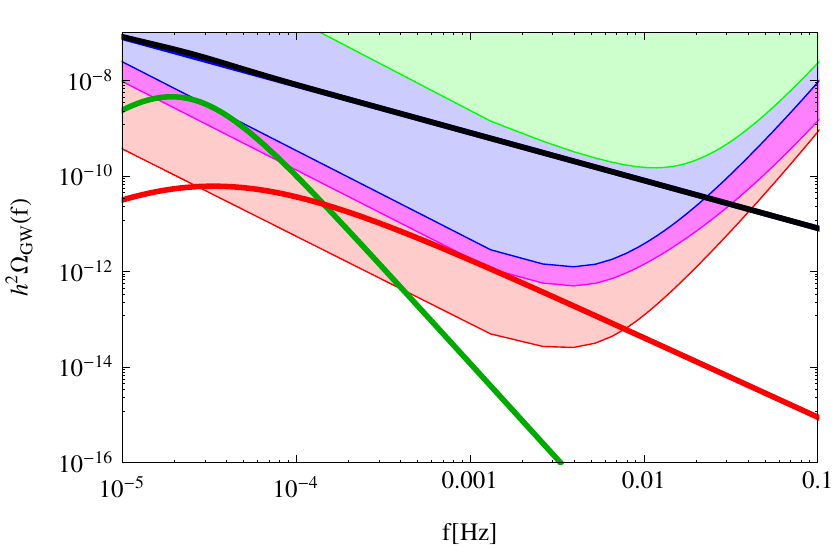}
\includegraphics[width=0.4\textwidth]{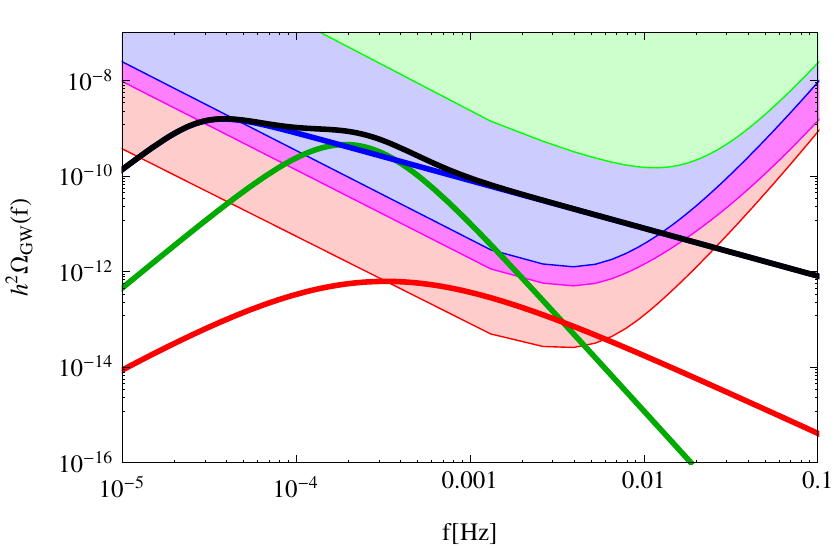}
\includegraphics[width=0.4\textwidth]{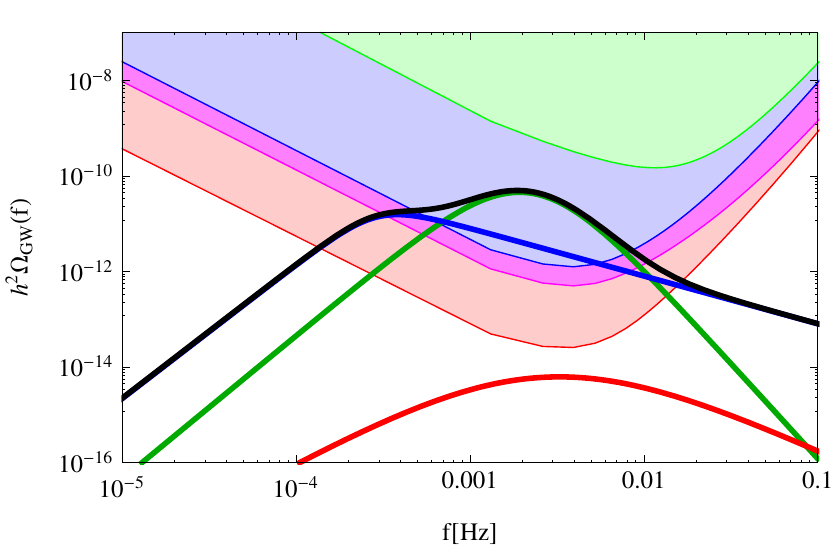}
\includegraphics[width=0.4\textwidth]{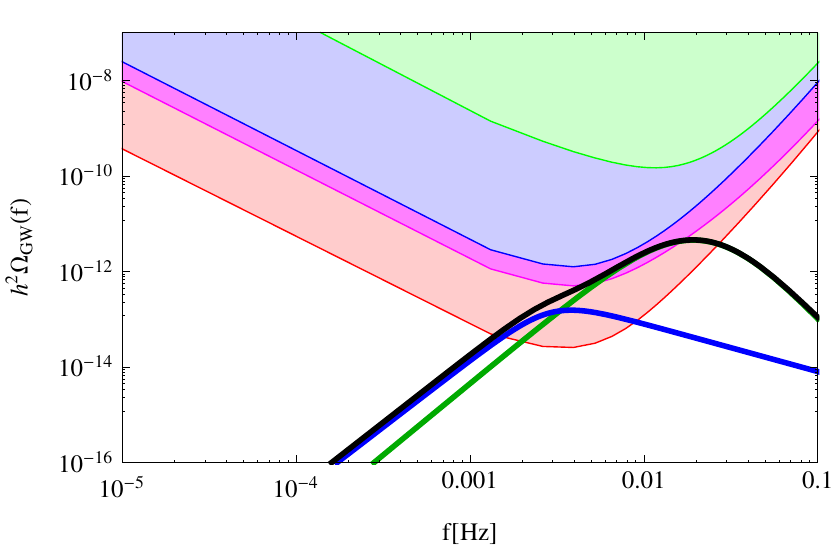}
\end{center}
\caption{\small \label{fig:spec_case2} Example of GW spectra in Case 2, for fixed $T_*=100$ GeV, $\alpha=1$, $v_w=1$, $\alpha_\infty=0.3$, and varying $\beta/H_*$: from left to right, $\beta/H_*=1$ and $\beta/H_*=10$ (top), $\beta/H_*=100$ and $\beta/H_*=1000$ (bottom). The black line denotes the total GW spectrum, the blue line the contribution from the scalar field, the green line the contribution from sound waves, the red line the contribution from MHD turbulence. The shaded areas represent the regions detectable by the C1 (red), C2 (magenta), C3 (blue) and C4 (green) configurations.}
\end{figure}

If the PT proceeds through runaway bubbles in a plasma (Case 2 above), the gradients of the scalar field also act as a source of GW together with sound waves and MHD turbulence. The corresponding examples of GW spectra are shown in Figure~\ref{fig:spec_case2}, for fixed $T_*$, $\alpha$, $v_w$ and $\alpha_\infty$, and varying $\beta/H_*$. Note that fixing $\alpha_\infty$ sets the relative amplitude of the scalar field-related and the fluid-related contributions. For small $\beta/H_*$ the contribution from the scalar field can dominate the GW spectrum, since the $\beta/H_*$ enhancement of the amplitude that operates for long-lasting sources is less relevant (c.f.~Eqs.~\eqref{eq:OmGsound} and \eqref{eq:Omenv}). As $\beta/H_*$ increases, the sound wave contribution gains importance (provided that $\alpha_\infty$ is large enough). At sufficiently high frequencies however the scalar field contribution always dominates because of its shallow decay: $p=1$ as opposed to $p=4$ and $p=5/3$, see Eqs.~\eqref{Senv},~\eqref{Ssw} and~\eqref{Sturb}. 

It is apparent that the total GW spectrum arising from a first-order PT depends on the interplay among the contributions of the different sources, which in turn are determined by the specific dynamics of the PT. On the one hand this is encouraging, since it opens up the possibility of investigating the dynamics of the PT. On the other hand, this is probably feasible only in the most optimistic PT scenarios and for the best eLISA configurations. Note that the highest GW signals are expected for runaway bubbles in vacuum (Case 3 above) for which the GW spectrum has the simplest shape, being determined only by the scalar field contribution.  
 
\subsection{Sensitivity to a First-Order Phase Transition}

With the eLISA sensitivity to a stochastic GW background
determined, we would like to assess eLISA's ability to detect GWs from primordial first-order PTs 
in a way that is as model-independent as possible. We have shown in the previous section that the predictions of the GW 
spectra differ in the three cases described in Sections 
\ref{sec:case1}-\ref{sec:case3}: we must therefore consider them separately. The 
most straightforward is Case 3, runaway bubbles in vacuum, since the GW 
spectrum depends only on the parameters $\beta/H_*$ and $T_*$. If plasma effects are important, 
the GW signal depends in addition on $\alpha$, $v_w$ and $\epsilon$, 
and moreover on $\alpha_\infty$ for the case of runaway bubble walls at 
finite $\alpha$. 

\begin{figure}
\begin{center}
\includegraphics[width=0.49\textwidth]{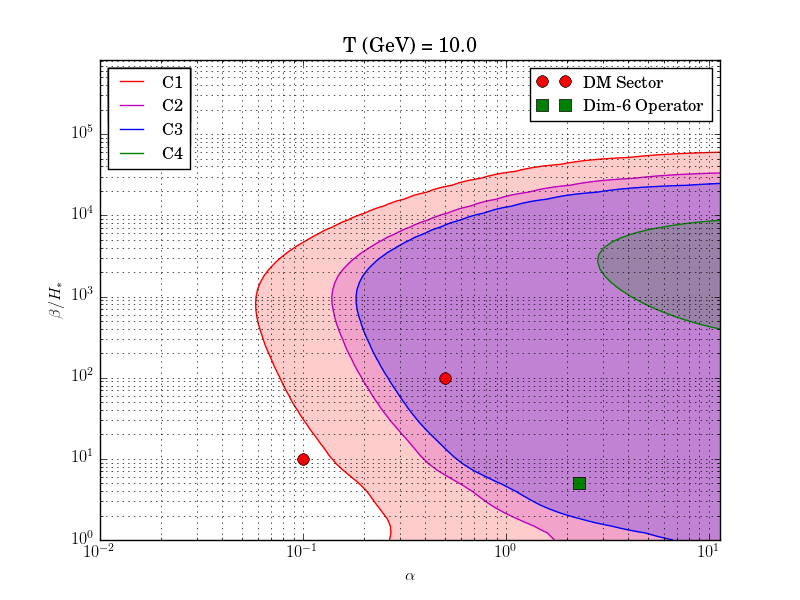}
\includegraphics[width=0.49\textwidth]{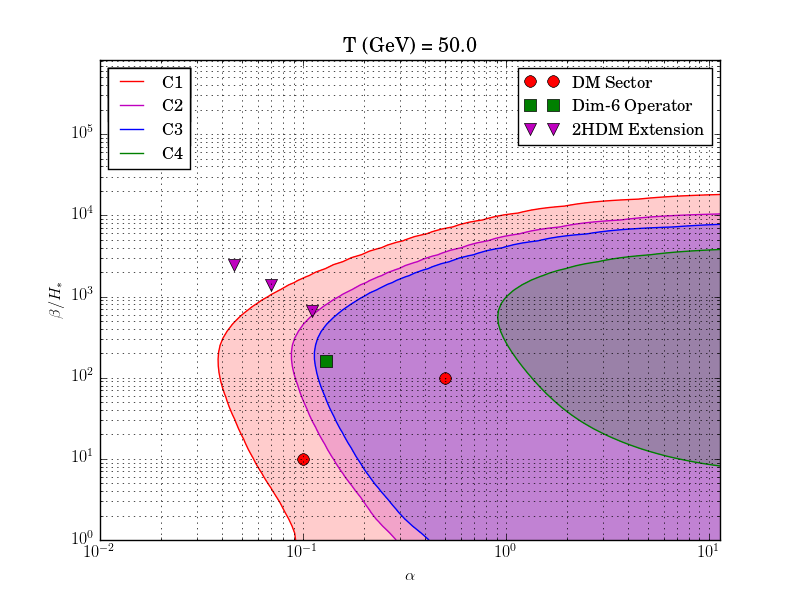}
\includegraphics[width=0.49\textwidth]{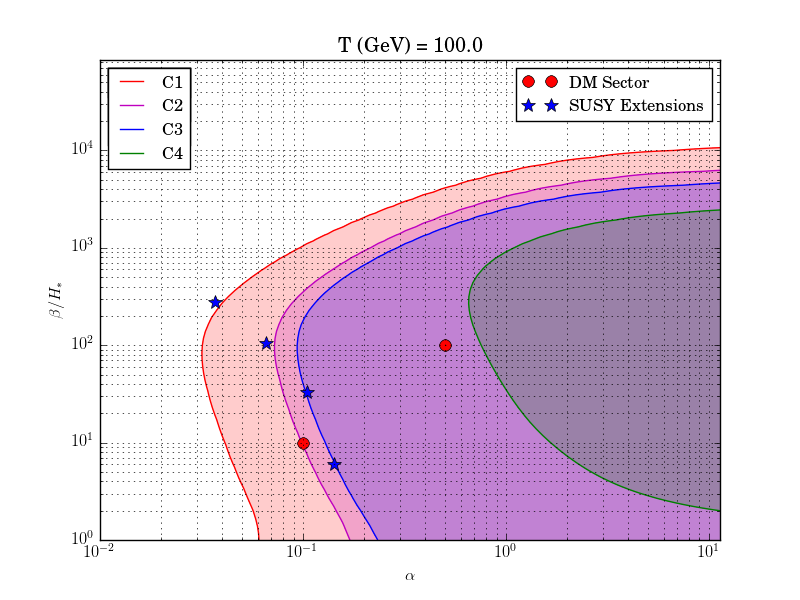}
\includegraphics[width=0.49\textwidth]{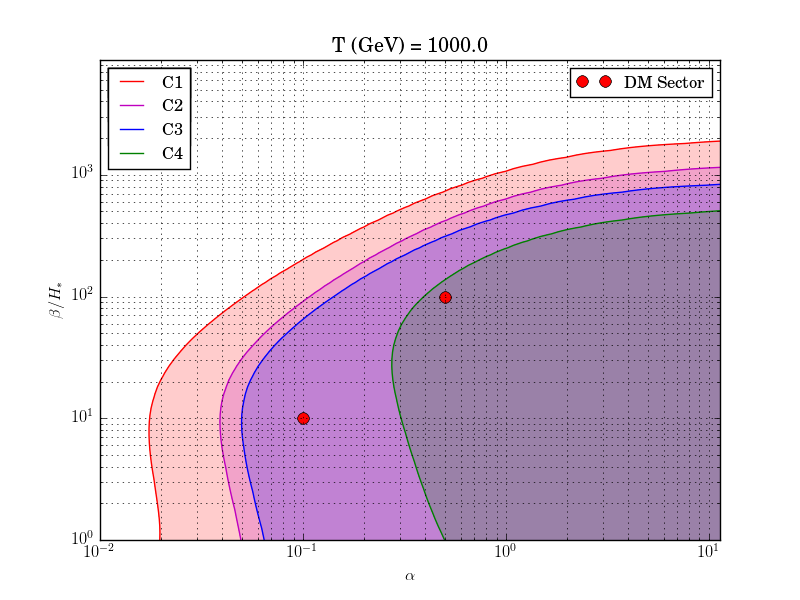}
\end{center}
\caption{\small \label{fig:case1} Projected eLISA sensitivity to Case 1: non-runaway relativistic bubble walls. 
Results are displayed for four  values of $T_*$ (indicated) and the four eLISA configurations described in Table \ref{tab:configurations}.
  The detectable region is shaded. Also 
  shown are benchmarks from various specific models, discussed in 
  Section~\ref{sec:models}. All other parameters are as described in the text.
  Note that the values of $T_*$ chosen correspond only
  approximately to the precise values for the benchmark points. The GW signal is given primarily
  by the contribution of sound waves (turbulence is negligible for the
  chosen value of $\epsilon$). }
\end{figure}

\begin{figure}
\begin{center}
\includegraphics[width=0.49\textwidth]{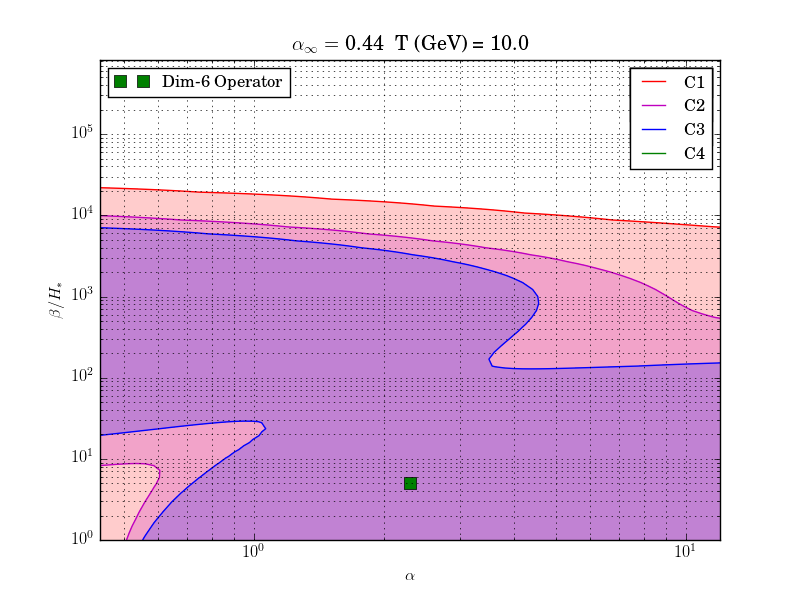}
\includegraphics[width=0.49\textwidth]{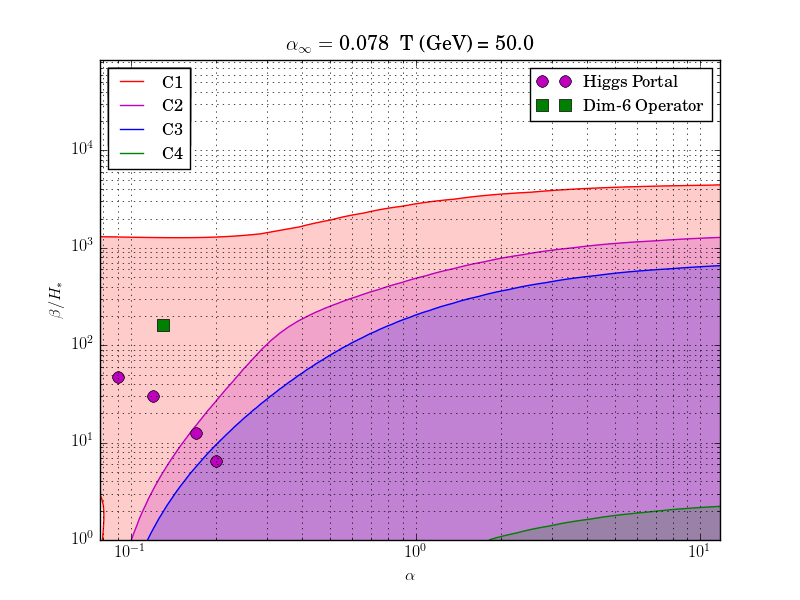}
\includegraphics[width=0.49\textwidth]{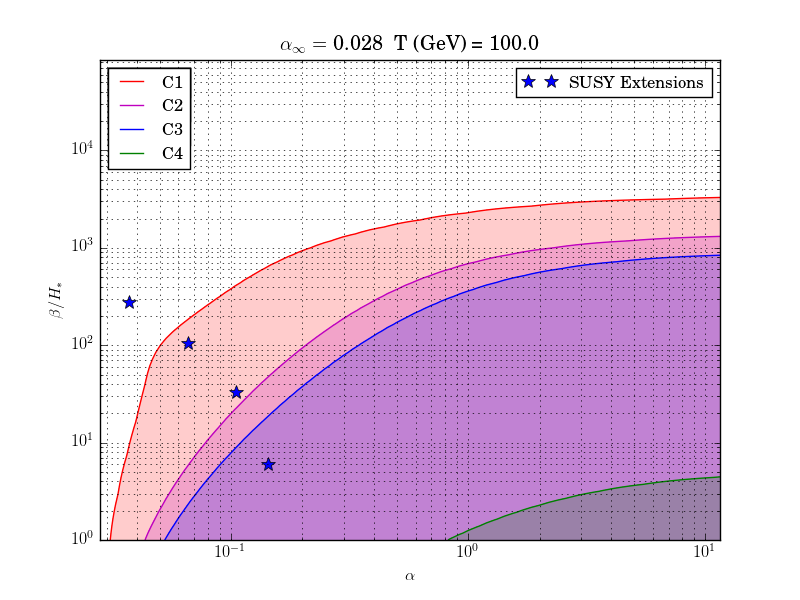}
\includegraphics[width=0.49\textwidth]{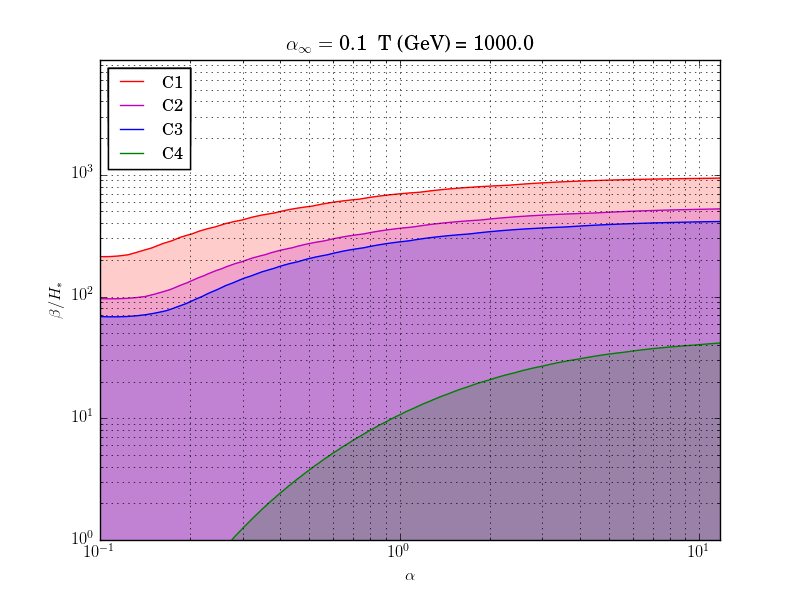}
\caption{\small \label{fig:case2} Projected eLISA sensitivity to Case 2: runaway bubble walls with
  finite $\alpha$. 
  Results
  are displayed for four  values of $T_*$ and $\alpha_\infty$ (indicated) and the four eLISA configurations described in Table \ref{tab:configurations}.
  The detectable region is shaded. Also shown are benchmarks from various specific models, discussed in 
  Section~\ref{sec:models}. All other parameters are as described in the text. 
   Note that the values of $T_*$ and $\alpha_\infty$ chosen correspond only
  approximately to the precise values for the benchmark points (as described in the text). The GW signal is given primarily by the contribution of the scalar field and of the sound waves.}
\end{center}
\end{figure}

\begin{figure}
\begin{center}
\includegraphics[width=0.49\textwidth]{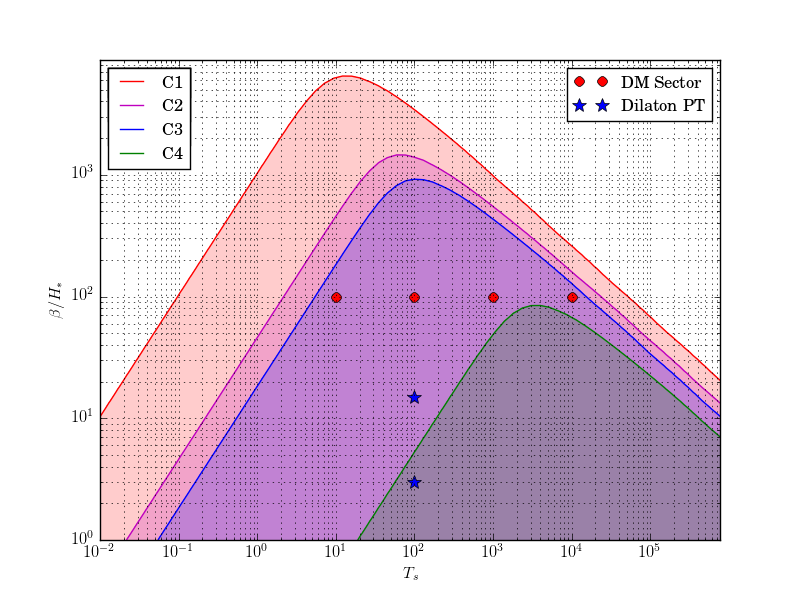}
\caption{\small \label{fig:case3} Projected eLISA sensitivity to Case 3: runaway bubble walls in vacuum. The region detectable by each configuration (c.f.~Table~\ref{tab:configurations}) is shaded. Also pictured are the predictions corresponding to the benchmark points discussed in Section~\ref{sec:models} that fall under Case 3.} 
\end{center}
\end{figure}

Once a model predicting a strong first-order PT is
chosen, these parameters are fixed. However, even without choosing a
model, general considerations prevent these quantities from varying
completely independently. For example, $\alpha$, $\beta/H_*$ and $T_*$
are related: as $\alpha$ increases, $\beta/H_*$ and $T_*$
typically decrease. This behaviour is due to the shape of the
Euclidean action of the tunneling solution, $S(T)$, as a function of
temperature~\cite{Huber:2007vva}; therefore, we can expect it to hold
quite generally, and to occur in many different models predicting a
strong first-order PT. However, the relationship between $\alpha$,
$\beta/H_*$ and $T_*$ cannot be specified analytically or numerically
in more precise terms without knowing $S(T)$ in detail i.e. without
restricting oneself to a given model. Therefore, it is not possible to
perform a model-independent analysis that realistically accounts for
the relation among $\alpha$, $\beta/H_*$ and $T_*$. As a consequence,
in the following we have chosen to let them vary freely~\footnote{A
  preliminary analysis of this type was carried out in
  \cite{Grojean:2006bp} under the simplifying but arbitrary assumption
  of a Jouguet detonation, which allows to express the bubble wall
  velocity as a function of $\alpha$.}, and in Figures~\ref{fig:case1},
\ref{fig:case2} we present contour plots in the plane
$(\alpha\,,~\beta/H_*)$ for different values of $T_*$ and
representative choices of $v_w$, $\epsilon$ and $\alpha_\infty$. In
Figure~\ref{fig:case3} the sensitivity is instead shown in the
$(T_*\,,~\beta/H_*)$ plane. It is important to keep in mind that not
all the points of the detection regions shown in
Figures~\ref{fig:case1}-\ref{fig:case3} can be realized
within a realistic model. See \cite{Binetruy:2012ze} for a very
similar analysis in the 2012 eLISA configuration (also called NGO at
the time).

Figure~\ref{fig:case1} shows the regions in $(\alpha\,,~\beta/H_*)$
which are accessible by the four eLISA configurations described in Section~\ref{sec:decthr}
for several values of $T_*$ in the case of non-runaway bubble walls
(taking $v_w=0.95$). We have set $\epsilon=0.05$. 
In all figures, the regions detectable by eLISA for each
eLISA configuration are shaded. As anticipated, {\it the two six-link configurations provide the best reach.}

The behaviour of the curves in Figure~\ref{fig:case1} can be understood as follows. Larger values of $\alpha$ and lower values of $\beta/H_*$
are easier for eLISA to detect because the amplitude of the GW spectrum
increases with $\alpha$ and decreases with $\beta/H_*$. At fixed $T_*$ and small $\beta/H_*$ only the high frequency tail of the spectrum can enter the sensitivity curve: as $\alpha$ increases, this happens at correspondingly smaller values of $\beta/H_*$. On the other hand, as $\beta/H_*$ increases for a fixed value of $\alpha$, the signal peak enters the detectable region, and finally exits it again for very high  $\beta/H_*$. The contours
flatten out at high $\alpha$, as the $\alpha$-dependence drops out from
the GW spectrum amplitude for $\alpha \gg 1$. Increasing the temperature causes 
smaller
$\alpha$-values to enter the detection region, because the peak
frequency of the spectrum is shifted towards higher frequencies where
the eLISA sensitivity is better. However, detection requires $\beta/H_*$ to be small
enough: as we have seen, increasing $\beta/H_*$ at fixed $T_*$ causes the peak
frequency to shift beyond the frequency window detectable by eLISA.

In Figure~\ref{fig:case1} we also show the benchmarks from various specific models, discussed in Section~\ref{sec:models}. The values of $T_*$ in each panel are chosen so as to approximately match the PT temperatures of the benchmarks.

Figure~\ref{fig:case2} applies to runaway bubbles (i.e.~$v_w=1$)
at finite $\alpha$. Again we take
$\epsilon=0.05$ and show several values of $T_*$ tuned to match the benchmarks of the models presented in Section~\ref{sec:models}. For concreteness, we have to fix the value of $\alpha_\infty$ in each panel: it is apparent that the contours of the detectable regions depends strongly on $\alpha_\infty$. In Appendix \ref{app:alphainfty} we provide extra figures showing the variation of the contours with $\alpha_\infty$. Each benchmark point of Figure~\ref{fig:case2} represents a model with a given value of $\alpha_\infty$ according to Eq.~\eqref{eq:alinf} (c.f.~the tables in Section~\ref{sec:models}). Consequently, for each panel we have chosen a value of $\alpha_\infty$ which is representative of all benchmarks, following the criterion that the position of the benchmark points with respect to the contours remains as unaltered as possible. For $T_*=1$ TeV there are no benchmarks: we have fixed $\alpha_\infty=0.1$. Note that in the four panels we only
plot the region $\alpha\geq \alpha_\infty$, for which the scalar field plays a role. For an explanation of the behaviour of the curves in Figure~\ref{fig:case2} we refer the reader to  Appendix \ref{app:alphainfty}.

We reiterate that the sound wave contributions in Figs.~\ref{fig:spec_case1}--\ref{fig:case2} rely on extrapolating the results found by simulations for $\alpha\lesssim 0.1$ and small fluid velocities to stronger transitions. The validity of this extrapolation remains to be determined. The reader should bear this in mind when interpreting our results, especially in cases where the sound wave contribution dominates for $\alpha\gtrsim 0.1$ (e.g. Case 1).

Finally, the projected eLISA sensitivity for Case 3 (runaway bubbles in vacuum) is shown in
Figure~\ref{fig:case3}. This case is parametrically simpler than the other two.
The GW signal, due entirely to the scalar field contribution, only depends on $T_*$
and $\beta/H_*$. eLISA's sensitivity to the stochastic background peaks 
for frequencies around $1-10$ mHz. Therefore, as the temperature
increases, the GW peak frequency $f_{\rm env}$ approaches this
region from below, and higher and higher values of $\beta/H_*$ are
detectable. However at some point the predicted peak frequency increases above
 the optimal eLISA value, and larger
amplitudes are necessary for detection; the
detectable values of $\beta/H_*$ start to decrease accordingly.

\subsection{Summary of Model-Independent Projections}
The model-independent analysis of this section shows that {\it the
  six-link configurations provide the most coverage to first-order cosmological 
  phase transitions. The configuration with four links and 2 million km
  arm length is, however, not much worse than that with six links and
  1 million km arm length.} Note that much better knowledge of the
instrumental noise and astrophysical backgrounds would be needed to
use the four-link configurations, since one cannot cross-correlate the
signal of the two effective, coincident detectors that the six-link
configurations provide. This is accounted for in the above analysis
through the technique explained in Section~\ref{sec:decthr}, and in
particular through the increase of ${\rm SNR}_{\rm thr}$. 
We stress that {\it our comparison between the four- and six-link configurations would change substantially if the
  analysis of Ref.~\cite{Adams:2013qma} were found to be unfeasible in practice, and/or the assumed prior knowledge of the noise 
  were unachievable.}

We now move on to consider how specific models map on to the
model-independent parameter space we have been discussing so far.

\section{\sc Testing Specific Models With Strong Phase Transitions} 
\label{sec:models}

Many scenarios beyond the Standard Model of particle physics predict first-order cosmological PTs.

Such a transition could have occurred during electroweak symmetry breaking.
Provided the Universe reached sufficiently high temperatures after inflation, 
electroweak gauge symmetry was likely unbroken after reheating. As the Universe 
cooled, the symmetry would then have been spontaneously broken when a Higgs 
field, charged under $SU(2)_L\times U(1)_Y$, acquired a VEV. In the Standard Model of particle physics, and for the observed value
of the Higgs mass, no transition occurs; electroweak symmetry is
instead broken at a cross-over~\cite{Kajantie:1995kf,
  Kajantie:1996mn}~\footnote{A GW spectrum is nonetheless generated by
equilibrium phenomena in the electroweak plasma, such as scatterings
between thermal constituents and collective phenomena. The resulting
spectrum \cite{Ghiglieri:2015nfa} is however tens of orders of
magnitude below eLISA sensitivity.}. However, many
well-motivated extensions of the Standard Model predict a strong
first-order electroweak PT instead of a cross-over. Such
a transition could have played a role in generating the observed
baryon asymmetry through the mechanism of electroweak
baryogenesis~\cite{Morrissey:2012db} and produced a spectrum of
gravitational radiation observable by eLISA, through the mechanisms
described above in Section~\ref{sec:signal}~\footnote{Typically, transport-driven electroweak baryogenesis requires sub-sonic bubble wall velocities, corresponding to small GW signals. There are exceptions however; see e.g. Ref.~\cite{Caprini:2011uz,No:2011fi}.}. 

Alternatively, there are various
extensions of the SM that predict strong first-order cosmological PTs not tied to the electroweak scale (or baryogenesis). Models solving the hierarchy problem via warped extra dimensions and dark matter sectors with non-trivial gauge structure are two such scenarios.

In the remainder of this section, we discuss several specific examples of models that predict strong first-order cosmological PTs, at the electroweak scale (Section~\ref{sec:EWPT_models}) or beyond (Section~\ref{sec:beyond}), with GW signals detectable by eLISA. Before proceeding, however, we briefly outline different mechanisms for generating strong PTs in the early Universe.

\subsection{Mechanisms for Generating a Strong First-Order Phase Transition}
There are a variety of well-studied mechanisms for generating a strongly first-order PT in scenarios beyond the Standard Model. 
Thermal loops of new bosonic modes can contribute a large cubic term to the finite temperature effective potential. This occurs for example in the 
minimal supersymmetric extension of the Standard Model (MSSM) with a light scalar top quark (stop). Even if the new degrees of freedom do not contribute 
substantially to the finite-temperature cubic term, it is also possible to rely primarily on the SM gauge boson contributions for the barrier, provided the 
free-energy difference between vacua at zero temperature is small enough (this in turn depends on the interplay of various terms in the potential), 
as may occur in extended Higgs sectors. 
Alternatively, new tree-level terms in the scalar potential can directly source a barrier (which can persist to zero temperature). This can occur either through 
new renormalizable or non-renormalizable operators.  In the former case, even if no cubic terms are present in the potential, a tree-level barrier can arise at 
finite temperature if one of the scalar fields obtains a non-vanishing VEV in an earlier (not necessarily first-order) PT.  

The aforementioned mechanisms typically rely on polynomial potentials, in which case the critical 
temperature, the nucleation temperature, the VEV of the scalar field at the 
minimum of the potential and  the position of the barrier in field space all typically share a common scale. 
An alternative and intriguing possibility arises in models with a dilaton-like potential that is nearly conformal. The potential can be described by a scale invariant 
function modulated by a slow evolution across scales, similar to the Coleman-Weinberg 
mechanism in which a slow Renormalisation Group evolution of the potential 
parameters can generate widely separated scales for the various aforementioned quantities.  The potential in this case is very shallow, and 
the position of the barrier  and the minimum of the potential can be very far 
apart.
As a result, a significant amount of supercooling and therefore a strong first-order PT can be obtained without a substantial tuning of parameters.

In the remainder of this section, we review some well-motivated scenarios
for physics beyond the Standard Model that can feature strong
first-order PTs via (combinations of) the mechanisms mentioned above. Our aim is to determine
reasonable values of $\alpha$, $\beta/H_*$ and $T_*$ that can actually
arise in realistic particle physics models.  For each model, we provide
some benchmark points (shown in Figures~\ref{fig:case1}-\ref{fig:case3}) and 
compute the parameters relevant for
predicting the spectrum of gravitational radiation. We also comment on
the detectability by eLISA in each case.

\subsection{Strong Phase Transitions at the Electroweak Scale}
\label{sec:EWPT_models}

\subsubsection{Supersymmetric Extensions of the Standard Model}
\label{sec:susy}

 Supersymmetric theories with
low-energy soft-breaking terms are theoretically well-motivated models featuring new scalar fields that can
naturally lead to strong electroweak PTs and, in turn, to sizeable
GW signals. 

The most well-studied supersymmetric extension of the SM is the MSSM. This theory
can give rise to a  
reasonably strong electroweak PT, provided the lightest
scalar top quark has a mass below that of the top quark~\cite{Carena,Delepine}.  This rough upper 
bound decreases substantially when the Higgs boson mass
is fixed at around 125\,GeV~\cite{mssm08,MikkoKari}, pushing the
model into severe tension with  LHC
stop searches and Higgs rate measurements~\cite{Menon:2009mz,
  Cohen:2012zza,Curtin:2012aa,Carena3}. At the same time,
independent of the PT, the MSSM is losing its original appeal
due to the large amount of fine-tuning required to obtain a Higgs mass of 
125 GeV. Both tensions can be alleviated in non-minimal supersymmetric models.

In singlet extensions of the MSSM, new $F$-term contributions can
increase the tree-level Higgs mass and, consequently, reduce the fine
tuning of the electroweak sector (compared to to the MSSM). Moreover, the presence of the scalar singlet enriches the Higgs
sector and has important consequences for the electroweak PT. Even after
imposing discrete symmetries on the field interactions, the (reduced)
parameter space still allows for strong PTs~\cite{Pietroni:1992in,
  Davies:1996qn,Apreda:2001us} while complying with all Run-I LHC
bounds~\cite{Kozaczuk:2013fga, Huang:2014ifa, Kozaczuk:2014kva}. In
particular, in the regime where the electroweak PT occurs in two steps
(i.e.~with the singlet acquiring a VEV first, and electroweak symmetry breaking occurring at a subsequent transition),
 very strong electroweak PTs seem likely~\cite{GWnmssm}.

For instance, in the four benchmark points of the singlet extension of
the MSSM studied in Ref.~\cite{GWnmssm}, the tree-level barrier
between the minima along the singlet and the SM-like Higgs directions
of the potential efficiently strengthens the electroweak PT. The
properties of the PTs corresponding to these four points were obtained and studied
in Ref.~\cite{GWnmssm}; the results are quoted in
Table~\ref{table:nmssm}. Notably, all points fulfill the runaway
requirement. Relativistic wall velocities are thus expected, although
obstructions could prevent the bubble from actually reaching the runaway
regime. Since the existence of these obstructions remains to be studied in this model, 
we consider the GW signal in both the
runaway and the non-runaway scenarios (Case 1 and Case 2, in the language of
Section~\ref{sec:scenarios}). The prospects for detecting the GW background from these benchmark points 
at eLISA are shown in Figures~\ref{fig:case1}
and~\ref{fig:case2} (bottom left panels) where the values of $\alpha$ and
$\beta/H_*$ of Table~\ref{table:nmssm} are displayed (the approximation 
$T_*\simeq 100\,$GeV has been
used). We find that 
(cf.~Figure~\ref{fig:case1}) {\it only the most sensitive eLISA configuration can probe the electroweak PT of the majority of the considered
benchmark points}. {\it For the relativistic non-runaway scenario, the C2 (six-link) configuration can probe two benchmark points, while if the
runaway scenario is realized} (cf.~Figure~\ref{fig:case2}) {\it it can only probe one of them}.

Other extensions of the MSSM beyond that considered in Ref.~\cite{GWnmssm} are possible. 
 However, such models seem unlikely to predict much stronger signals than those considered here, at least for electroweak (i.e.~$SU(2)_L\times U(1)_Y$--breaking)
transitions and without significantly departing from minimality. The predictions could differ substantially when
considering PTs occurring before the electroweak transition. Further study
is required to clarify this possibility.

\begin{table}
\begin{center}
\begin{tabular}{ | c || c | c | c | c | }
  \hline                       
      & A & B & C & D \\
  \hline
  \hline
 $T_{*}$ [GeV]	& $112.3$ 	& $94.7$	& $82.5$ & $76.4$ \\
 $\alpha$ 		& $0.037$ 	& $0.066$	& $0.105$ & $0.143$ \\
 $\beta/H_*$ 		& $277$ 	& $105.9$ 	& $33.2$ & $6.0$ \\
$\phi_*/T_*$        & $1.89$	& $2.40$	& $2.83$ & $3.12$\\ 
\hline
\end{tabular}
\caption{\small
\label{table:nmssm}
Characteristics of the electroweak PT predicted for the benchmark points of the singlet extension of the MSSM analyzed
in Ref.~\cite{GWnmssm} (see Section~\ref{sec:susy}).}
\end{center}
\end{table}

\subsubsection{The Standard Model with Additional Scalars}
\label{sec:Hportal}

Another simple class of scenarios that can give rise to observable GWs are (non-supersymmetric) extensions of the Standard Model scalar sector. 
The new scalar(s) can either be singlets or charged under the Standard Model gauge groups. 
 
New SM gauge singlet scalar fields can couple directly to the Higgs field via renormalizable operators. Such models are 
attractive from the standpoint of dark matter~\cite{Craig:2014lda}, baryogenesis~\cite{Profumo:2014opa}, and solutions to the hierarchy 
problem without colored top partners~\cite{Craig:2013xia}. New terms in the tree-level potential can allow 
for strong electroweak PTs and an observable level of gravitational radiation.

To see what this entails, let us consider the Higgs portal scenario with a real 
gauge singlet scalar field, which is the simplest case and illustrates some key features of this 
scenario. We will restrict ourselves to the case where the new gauge singlet scalar, $S$, is charged under a discrete $\mathbb{Z}_2$ symmetry and has no 
VEV at zero temperature~\cite{Espinosa:2011ax}. All other Standard Model particles are assumed to transform trivially under the 
new $\mathbb{Z}_2$ symmetry. The most general renormalizable scalar potential in this case can be 
written as 
\begin{equation}
\label{eq:singlet_pot}
V(H,S)=-\mu^2(H^{\dagger}H)+\lambda (H^{\dagger}H)^2+\frac{1}{2}a_2 (H^{\dagger}H)S^2+\frac{1}{2}b_2 
S^2+\frac{1}{4} b_4 S^4 \, .
\end{equation}
Here, the CP-even neutral component of $H$ is identified with the 125 GeV Standard Model--like Higgs. The discrete symmetry ensures that $S$ is stable. 
The singlet can thus contribute to the observed dark matter relic abundance. Note that $S$ can also be a component of a scalar charged under other symmetries 
(e.g.~a hidden sector gauge group).
One should also bear in mind that this model resides in a subspace of a larger model parameter space without the $\mathbb{Z}_2$ symmetry. Going beyond 
the $\mathbb{Z}_2$ limit opens up additional parameter space for a strong first-order PT. 

This ``Higgs portal'' model can give rise to a strong electroweak PT in primarily two ways. If the parameter $b_2<0$, the singlet can be 
destabilized from the origin at finite temperature (along the $S$ direction in field space). The $H^{\dagger}H S^2$ term then provides a cubic term to 
the effective potential, and hence can contribute to a barrier along the direction connecting the $\langle S\rangle \neq 0$ and electroweak 
vacua (in which $\langle S\rangle = 0$). This is another example of a two-step transition. Alternatively, if $b_2>0$, the singlet will be stabilized at 
the origin at all temperatures. Nevertheless, large zero-temperature loop effects can lower the SM-like Higgs quartic coupling to increase the value of $\phi(T_n)$ inside the bubble.  Finite-temperature loop effects can also play a role (alongside the SM gauge, Higgs, and Goldstone bosons) in contributing to a 
barrier between the origin and the electroweak vacuum in this case.

To illustrate the characteristics of the electroweak PT in this scenario, we consider four benchmark points. These points, 
labeled A--D, are chosen such that $m_S=250$ GeV, and correspond to $(a_2,b_4) = (2.8, 2.1)$, $(2.9, 2.6)$, $(3.0, 3.3)$, and $(3.1,  4.0)$, respectively; they satisfy all current phenomenological constraints and are particularly difficult to test at colliders~\cite{Curtin:2014jma}. The corresponding PT 
parameters are displayed in Table~\ref{table:scalars}; they are obtained by using the full 1-loop finite-temperature effective potential 
and including the resummation of daisy diagrams. For such strong transitions, the bubble wall is expected to run away without obstruction~\cite{Kozaczuk:2015owa}, corresponding 
to Case 2 of Section~\ref{sec:scenarios}. The sensitivity of eLISA to these points is shown in Figure~\ref{fig:case2}, top-right panel. We find that {\it the most sensitive eLISA configuration can probe all four benchmark points, while C2 (six-links) and C3 (four-links) can probe benchmark point D, with point C residing at the edge of the region detectable by C2)}. The observational situation is therefore similar to the one discussed in Section~\ref{sec:susy}.

\begin{table}
\begin{center}
\begin{tabular}{ | c || c | c | c | c | }
  \hline                       
      & A & B & C & D \\
  \hline
  \hline
 $T_{*}$ [GeV]	& $70.6$ 	& $65.2$	& $59.6$ & $56.4$ \\
 $\alpha$ 		& $0.09$ 	& $0.12$	& $0.17$ & $0.20$ \\
 $\beta/H_*$ 		& $47.35$ 	& $29.96$ 	& $12.54$ & $6.42$ \\
$\phi_*/T_*$        & $3.39$	& $3.70$	& $4.07$ & $4.32$\\ 
\hline
\end{tabular}
\caption{\small
\label{table:scalars}
Characteristics of the electroweak PT predicted for the Higgs portal benchmark points
discussed in Section 
\ref{sec:Hportal}.}
\end{center}
\end{table}

While the benchmarks shown above feature thermal transitions (i.e. the $O(3)$-symmetric bounce minimizes the four-dimensional 
Euclidean action at $T_n$), it should be noted that this model can allow for a metastable 
electroweak-symmetric phase to persist to zero temperature. This
suggests that very strongly supercooled transitions occurring in vacuum (Case 3) 
may be possible. The resulting GW 
signals in this case have not been previously analyzed, but would be worthwhile 
to consider in future work.

\vspace{3mm} 
 
New scalars charged under the SM gauge group are also phenomenologically and theoretically well-motivated.  
When transforming nontrivially under the electroweak gauge group, such scalars can participate in electroweak symmetry breaking, with a 
potentially significant impact on the electroweak PT. The simplest realization of this scenario is that of a two-Higgs-doublet-model (2HDM), in which the 
SM Higgs sector is enlarged by a second scalar doublet. These scenarios can accommodate electroweak baryogenesis~\cite{Fromme:2006cm} and can result in the generation of GWs at the electroweak PT (with preliminary studies 
having been carried out in~\cite{Kakizaki:2015wua}). The scalar potential is~\footnote{We assume for simplicity CP conservation,
as well as a $\mathbb{Z}_2$ symmetry (softly broken by $\mu^2$ in Eq.~\eqref{2HDM_potential}) for phenomenological reasons, namely the absence of 
flavour-changing neutral currents in the Higgs sector.}
\begin{eqnarray}	
\label{2HDM_potential}
V(H_1,H_2) &= &\mu^2_1 \left|H_1\right|^2 + \mu^2_2\left|H_2\right|^2 - \mu^2\left[H_1^{\dagger}H_2+\mathrm{h.c.}\right] + \frac{\lambda_1}{2}\left|H_1\right|^4 +\frac{\lambda_2}{2}\left|H_2\right|^4\nonumber \\
&+& \lambda_3 \left|H_1\right|^2\left|H_2\right|^2 + \lambda_4 \left|H_1^{\dagger}H_2\right|^2+ \frac{\lambda_5}{2}\left[\left(H_1^{\dagger}H_2\right)^2+\mathrm{h.c.}\right]\, .
\end{eqnarray}
After electroweak symmetry breaking, the presence of the two doublets $H_1$, $H_2$ yields three new physical states in addition to the 125 GeV Higgs $h$: 
a charged scalar $H^{\pm}$ and two neutral states $H_0$, $A_0$. 

In this class of scenarios, a strong electroweak PT is driven by a decrease in the free-energy difference between 
the electroweak-symmetric local maximum and electroweak-broken phase at $T=0$, w.r.t.~the SM prediction.  
The effect of this decrease on the strength of PT is in fact similar
to the reduction of the SM-like Higgs quartic coupling occurring for singlet ``Higgs-portal" scenarios with $b_2 > 0$, as described above. 
Moreover, such a decrease in the free energy difference between the origin (in field space) and the vacuum with broken electroweak symmetry is highly correlated
with a large mass splitting between the new states $A_0$, $H_0$~\cite{Dorsch:2013wja,Dorsch:2014qja}, which provides an appealing connection 
to LHC signatures.

For 2HDM scenarios, it is always true that $\alpha < \alpha_{\infty}$, and so these models belong to Case 1
of Section~\ref{sec:scenarios}. We present here three benchmark points studied in~\cite{GW2HDM} with increasing strength of the electroweak PT, labelled A--C in Table \ref{table:2HDM}. The eLISA sensitivity to these benchmarks is shown in Figure~\ref{fig:case1}, top-right panel. We find that {\it only one benchmark point lies in the region detectable by C1. This point may also be marginally detectable by C2. Four-link configurations provide no access to this model.}

\begin{table}
\begin{center}
\begin{tabular}{ | c || c | c | c | }
  \hline                       
      & A & B & C \\
  \hline
  \hline
 $T_{*}$ [GeV]	& $68.71$ 	& $61.25$	& $51.64$ \\
 $\alpha$ 		& $0.046$ 	& $0.070$	& $0.111$ \\
 $\beta/H_*$ 		& $2446$ 	& $1383$ 	& $663$ \\
$\phi_*/T_*$        & $3.15$	& $3.69$	& $4.53$ \\ 
\hline
\end{tabular}
\caption{\small
\label{table:2HDM}
Characteristics of the electroweak PT predicted for the benchmark points considered for the 2HDM in Section 
\ref{sec:Hportal}.}
\end{center}
\end{table}

Finally, let us also point out that new scalar fields can also be charged under the SM color group, $SU(3)_c$. Such a scalar could trigger a 
strong first-order color-breaking transition in the early universe~\cite{Patel:2013zla}. The broken gauge symmetry can be restored by a subsequent transition 
to the standard electroweak vacuum.  As argued in Ref.~\cite{Patel:2013zla}, color-breaking transitions in this two-step setup typically occur at temperatures 
close to the TeV scale. This may provide another interesting target for the eLISA experiment. We leave a study of the GW signatures of this scenario to future work.

\subsubsection{The Standard Model with Higher-dimensional Operators}
\label{sec:dimsix}

We have seen that tree level modifications of the Higgs potential can easily 
make the electroweak PT strongly
first-order. The models of the previous sections considered cases in which 
new terms in the potential involving new (relatively light)  scalar fields significantly strengthen the 
electroweak PT. Alternatively, the effects of \emph{heavy} new physics on the PT
can be studied and illustrated in a model-independent manner using an effective field 
theory approach, for instance by adding dimension-6 operators in the Higgs 
potential allowing for a negative quartic 
coupling~\cite{Grojean:2004xa,Delaunay:2007wb}:
\be
V(\phi)=\mu^2|\phi|^2-\lambda|\phi|^4+\frac{|\phi|^6}{\Lambda^2}
\label{eq:eff}~.
\ee
This model illustrates the typical correlations expected
between small values of $\beta/H_*$ and large $\alpha$ mentioned earlier.
Contours of $\alpha$ and $\beta/H_*$ were computed in the $(m_h,\Lambda)$--plane
for the complete one-loop finite temperature effective potential
associated with Eq.~(\ref{eq:eff}) in Ref.~\cite{Delaunay:2007wb}. The region corresponding to a sizable
GW signal from the electroweak PT is confined to values of $\Lambda$
below 1 TeV. The tension between such a low cutoff and the LHC bounds
remains to be investigated. Nevertheless, focusing on the GW signal, we
consider two benchmark points where $\Lambda$ is around 600\,GeV 
\cite{Huber:2007vva}, both in the relativistic, non-runaway case and in the 
runaway with finite $\alpha$ case (the predicted wall velocity
remains to be studied in such non-renormalizable models, although runaways seem likely, given the tree-level origin of the barrier between vacua). The resulting features of the electroweak PT 
are quoted in
Table~\ref{table:dim6}. We find that {\it for most benchmark points
  the GW signal can be detected by all eLISA configurations except 
  C4. However, only C1 can detect all benchmark points.}

\begin{table}
\begin{center}
\begin{tabular}{ | c || c | c |  }
  \hline                       
      & A & B \\
  \hline
  \hline
  $T_{*}$ [GeV]	        & 63 	& 26  \\
 $\alpha$ 		& 0.13 	& 2.3  \\
 $\beta/H_*$ 		& 160 	& 5  \\
$\phi_*/T_*$       & 4	& 9.5 \\
\hline
\end{tabular}
\caption{\small
\label{table:dim6}
Characteristics of the electroweak PT in the SM plus a dimension-6
effective operator for two benchmark points taken from Ref.~\cite{Huber:2007vva}:
$\Lambda\sim 600\,$GeV (A) and $\Lambda\sim 576\,$GeV (B), see Section 
\ref{sec:dimsix}.  }
\end{center}
\end{table}

\subsection{Strong Phase Transitions Beyond the Electroweak Scale} 
\label{sec:beyond}

As we have seen, observable GWs may have been produced
at the electroweak scale in various scenarios beyond the Standard Model. However, 
there are other scenarios for new physics that may have given rise to a strong first-order
cosmological PT in the early Universe. We discuss two such examples here.  

\subsubsection{Dilaton-like Potentials and Naturally Supercooled Transitions}
\label{sec:dilaton}

Models with a spontaneously broken (approximate) conformal symmetry feature a pseudo-Nambu-Goldstone boson 
associated with the broken symmetry; this field is known as the dilaton. 
The scalar potential $V_{\sigma}(\sigma) $ of the dilaton field, $\sigma$,
can be parametrized by a scale invariant function modulated by weakly scale-dependent function:
\be
V_{\sigma}(\sigma)= \sigma^4 \times P(\sigma^{\epsilon}) \ \ \mbox{where } \ 
|\epsilon| \ll 1
\label{eq:dilatonpotential}
\ee 
A particularly interesting and well-motivated class of scenarios
arises when the quadratic term for the Higgs field $\phi$ 
is controlled by the 
VEV of the dilaton $\sigma$:
\be
V(\sigma,\phi)= V_{\sigma}(\sigma) +\frac{\lambda}{4} (\phi^2-\xi \ \sigma^2)^2
\label{eq:higgsdilatonpotential}
\ee
where $\xi$ is a constant.
In particular, this potential is precisely that of the 5D Randall-Sundrum models 
\cite{Randall:1999ee}, which provide an elegant solution to the hierarchy problem of the SM. 

Assuming the Higgs is localized on the IR brane at a 
distance $y=r$ from the UV brane (localized at $y=0$), the 4D  effective action 
for the Higgs is
\be
{\cal L}_4=e^{-2k\pi r} \eta^{\mu\nu} D_{\mu}\tilde{H} D_{\nu} \tilde{H} 
-e^{-4k\pi r} \lambda(|\tilde{H}|^2-v_P^2)^2
=\eta^{\mu\nu} D_{\mu}{H} D_{\nu} {H} - \lambda \left(|{H}|^2-\frac{v_P^2}{k^2} 
\sigma^2 \right)^2
\ee
where $v_P \sim \Lambda_{UV}\sim m_{Pl}\sim k$, $H$ is the canonically 
normalized field $H=e^{-k \pi r} \tilde{H}$ and  the radion field is
\be
\sigma \equiv k e^{-k \pi r}.
\ee
We also define the scale
\be
 \Lambda_{IR} \equiv \langle \sigma \rangle 
\ee
which is generated once the radion is stabilized and is exponentially warped 
down from the Planck scale due to the Anti de Sitter (AdS) geometry. We also have
$\xi=v^2/\Lambda_{IR}^2$.
For the 5D AdS metric, the effective 4D potential for the radion was shown  to 
be dilaton-like (Eq.~\ref{eq:dilatonpotential}), 
independently of the inter-brane distance stabilization mechanism 
\cite{Goldberger:1999uk,Rattazzi:2000hs,Garriga:2002vf}. We therefore recover 
the scalar potential Eq.~(\ref{eq:higgsdilatonpotential}) for the coupled 
radion-Higgs system. Solving the weak/Planck scale hierarchy leads to $\Lambda_{IR} 
\sim {\cal O}$(TeV).

The cosmological implications of the potential $V_{\sigma}(\sigma)$ in Eq.~(\ref{eq:higgsdilatonpotential})
 were considered in Refs.~\cite{Nardini:2007me,
  Konstandin:2011dr}.  A very strong first-order PT
typically occurs for this type of potential.  In the first
investigation of the associated PT, it was argued that
the transition to the minimum of the radion potential could not complete
\cite{Creminelli:2001th}.  This conclusion essentially followed from a
thin-wall estimate of the critical bubble action and assuming that
tunneling would take place directly to the minimum of the potential.
The key point stressed in Ref.~\cite{Randall:2006py} is that the PT can actually complete through tunneling to a value of the
field much smaller than the value at the minimum of the potential and
subsequently rolling towards the minimum.  This is typical of
very shallow potentials.  As emphasized in
Ref.~\cite{Konstandin:2011dr}, the value of the field at tunneling,
$\sigma_r$, is $\sigma_r \sim \sqrt{ \sigma_{+ }\sigma_{-}} $ where
$\sigma_{+}$ and $\sigma_{-}=\Lambda_{IR}$ are the positions of the maximum and
minimum of the potential respectively.  The nucleation temperature
$T_n$ is proportional to $\sigma_r$ and given by
\cite{Konstandin:2011dr,Konstandin:2010cd}
 \be T_n \sim 0.1
\sqrt{\sigma_{+ }\sigma_{-}} \sim 0.1 \ \Lambda_{IR} \ \sqrt{\frac{\sigma_+
  }{\sigma_{-}}} ~.\ee
 For a standard polynomial potential, $\sigma_+
\sim \sigma_- \sim \sigma_r \sim T_n$. In contrast, for the very
shallow dilaton-like potential, $\sigma_+ \ll \sigma_-$, and the
nucleation temperature is parametrically much smaller than the scale
associated with the minimum of the potential. We therefore naturally
get a stage of supercooling before the PT completes. The
hierarchy between $\sigma_{-}$ and $\sigma_{+}$ can be as large as the
Planck scale/weak scale hierarchy: ${\sigma_{-}}/{\sigma_{+}} \lesssim
{\Lambda_{UV}}/{\Lambda_{IR}}$.  Therefore the nucleation temperature
can be as low as $T_n\sim 0.1
\Lambda_{IR}\sqrt{\Lambda_{IR}/\Lambda_{UV}}$~\cite{Konstandin:2011dr}.  We obtain $T_n\sim 35$
MeV if $ \Lambda_{IR}=5 $ TeV and $\Lambda_{UV}=M_{Pl}$, while
$T_n\sim 0.1$ GeV if $ \Lambda_{IR}=1 $ TeV and $\Lambda_{UV}=10^{10}$
GeV.  Note that at scales below $\Lambda_{\rm QCD}$, Eq.~(\ref{eq:dilatonpotential}) will be modified, 
since QCD breaks conformal invariance. 
While a delayed electroweak PT down to
the QCD scale is in principle a general outcome in this framework (and
interesting from the standpoint of e.g. cold electroweak baryogenesis with the QCD axion
\cite{Servant:2014bla}), this modification of the scalar potential
around the QCD scale, which we do not account for here, will affect the detailed predictions of this model.

We have argued that the nucleation temperature of the PT associated with a
potential of the form in Eq.~(\ref{eq:dilatonpotential}) can be many orders of magnitude
smaller than the scale given by the VEV at the minimum of the
potential.  In this case, plasma effects can be ignored and the GW
signal comes from runaway bubbles in vacuum. As discussed in Section~\ref{sec:case3}, while the PT temperature $T_n$ can in principle be 
much smaller than $\Lambda_{IR}\sim {\cal
  O}$(TeV), the temperature $T_*$ in the various GW formulae will be set by the reheating temperature, 
  $T_* \approx T_{\rm reh}$. This temperature is typically somewhat below $\Lambda_{IR}$ and the peak frequency (Eq.~(\ref{eq:envPeak})) can easily 
  fall within the range that will be probed by eLISA. In particular, for a dilaton VEV at the
TeV scale, as motivated by the Randall-Sundrum scenario, and more
generally by composite Higgs models which have a strongly coupled
sector at the TeV scale, the predicted GW signal can be detectable by eLISA \cite{Randall:2006py,Konstandin:2010cd}.

The prospects for detecting the gravitational wave signal predicted by such a scenario are analyzed for two
benchmark points, with PT properties summarized in
Table~\ref{table:dilaton}. The two benchmark points A and B represent
the parameter space region in which this model solves the hierarchy problem without introducing a significant ``little hierarchy'' between the EW scale and $\Lambda_{IR}$.
In computing $\beta/H_*$, Eq.~(\ref{eq:boH_vacuum}) is used, as is appropriate for vacuum transitions, since $T_*\approx T_{\rm reh}\gg T_n$.  
As shown in Figure~\ref{fig:case3}, {\it the prospects for observing GWs from this model are very good. 
Benchmark point A can be detected by all configurations while 
benchmark point B can be observed by all but C4}.

\begin{table}
\begin{center}
\begin{tabular}{ | c || c | c | c | c | c |}
  \hline                       
     & A & B\\
  \hline
  \hline
  $T_{*}$ [GeV]	         & 100 	& 100  \\
  $\beta/H_*$ 		& 3 	& 15	  \\
\hline
\end{tabular}
\caption{\small
\label{table:dilaton}
Characteristics of the PT predicted for the benchmark points of the 
dilaton-like scenario in Section \ref{sec:dilaton}.
}
\end{center}
\end{table}

Very strong first-order PTs associated with nearly conformal dynamics at the TeV scale
are interesting from the point of view of the hierarchy problem, but a similar situation may arise at a different scale. In this case, the discussion of this section remains applicable as far as the strength of the GW signal is concerned, although the predicted peak frequency will be affected. 

\subsubsection{First-Order Phase Transitions in a Dark Matter Sector}
\label{sec:DMS}

Models in which dark matter (DM) is a stable bound state of a confining dark sector 
are well motivated. Common examples are $SU(N)$ dark sectors with $n_f$ light 
dark quarks, where the DM candidate is a dark baryon-like 
state~\cite{Bai:2013xga}, as well as cases with no massless quarks, in which 
case dark glueballs are DM candidates~\cite{Boddy:2014yra}. The DM mass is 
typically of the order of the confinement scale, which in turn is set by the scale of the 
associated symmetry breaking PT. In a large class of models 
the PT is first order, and therefore can give rise to a GW signal. 

Viable DM models of this class have masses ranging from ${\cal O}(10)$~MeV up to 
${\cal O}(100)$~TeV. The GW signal falls into the eLISA frequency range for DM 
masses ranging from $10$~GeV to $10$~TeV. In particular the high mass range is 
difficult to test in current collider experiments, and GWs provide a unique 
window to probe some aspects of these models. 

The first-order nature of the PT in these models can be determined
from very basic symmetry arguments~\cite{Schwaller:2015tja}. For an
$SU(N)$ dark sector with $N \geq 3$ and $n_f$ light quarks the PT is
first order for $n_f = 0$ or $3 \leq n_f \lesssim 4 N$. The PTs occur in the non-perturbative regime of the theories, so
the details of the dynamics are currently not known. This can be
improved in the future using lattice simulations.

Some models might feature a rather weak PT, which will typically make them inaccessible by eLISA. On the other hand, studies of holographic
PTs~\cite{Creminelli:2001th,Randall:2006py,Nardini:2007me,
  Konstandin:2010cd, Konstandin:2011dr} discussed in Section
\ref{sec:dilaton} find cases with very strong transitions, and
therefore one can hope that models close to the conformal window
(i.e.~with $n_f$ close to $4N$) will also fall into this class. Thus, we expect at
least a subset of the models to feature a sufficiently strong
signal to be detectable by eLISA. 

We consider 
a scenario with relativistic (non-runaway) bubbles in a plasma (Case 1), and 
a scenario with runaway bubble walls in vacuum (Case 3). We do not consider the possibility of runaway bubbles in a plasma (Case 2), 
simply because the prospects for GW detection by eLISA in this case depend sensitively on $\alpha_{\infty}$ (c.f. Appendix~\ref{app:alphainfty}), which is
not known for these scenarios. In fact, as the present
understanding of the model does not allow for proper quantitative
estimates, we simply make some reasonable guesses for all of the relevant PT parameters. 
Specifically, in the first case we assume $T_*=10, 50, 10^2, 10^3$\,GeV. For
each temperature we consider $\beta/H_*=10$ as well as
$\beta/H_*=100$, respectively accompanied by $\alpha=0.1$ and
$\alpha=0.5$. In the second case we assume $T_*=10, 10^2, 10^3, 10^4$\,GeV and $\beta/H_*=100$. 
{\it We stress that these values
  are simply educated guesses, since our present understanding of the model does not
  allow for proper quantitative estimates}.  We
analyze the GW signal detectability for Case 1 and Case 3
in~Figures~\ref{fig:case1} and~\ref{fig:case3}. We find
that, {\it in Case 1, the C4 configuration can only probe one of the benchmark points, while C1 can detect all but one. C2 and C3 provide comparable sensitivity to these points. For Case 3, all but C4 can detect the benchmark points considered.}

Despite its numerous issues that require future clarification, the present
model deserves some attention as it highlights a link between eLISA and
DM experiments.
Dark matter model building in recent years has focussed on non-minimal
models with additional dynamics, and this model is a broad subclass of
this category. From the point of view of theoretical particle physics
there is a strong desire to find alternative ways to probe these
models, since they are relatively hidden from collider and direct DM
searches. In contrast, the GW signal does not depend on the interaction
strength between the dark and visible sectors, and thus provides a
unique way to probe these scenarios.
In particular, most of the parameter space will not be
excluded by other experiments within the next 20 years, and so eLISA may provide the first evidence
in favor of such a dark sector. The criteria for first-order PT are simple and
generic, and the models considered so far are minimal. Other classes
of models in which DM is associated with a first-order PT are
conceivable, and the models considered here could also be extended, for
example to enhance the strength of the transition. Future work considering these possibilities is
warranted.

\section{\sc Summary and Conclusions} \label{sec:conc}

We have seen that many scenarios beyond the SM predict
a GW signal from
first-order phase transitions in the early Universe that can be observed by eLISA.  
Although ongoing efforts at the LHC will be able to test
some of these scenarios, there remain many examples that cannot be probed through
collider experiments on a timescale comparable to that of eLISA, if at all.

The eLISA detector has the potential to provide us with valuable information about
electroweak-scale physics that cannot be obtained from high-energy colliders,
for example concerning the dynamics of the electroweak phase transition and the shape
of the Higgs potential, or
concerning hidden sectors and dark
matter at the weak scale. Violent processes in the early Universe may in fact have taken
place in a  sector with very feeble couplings to the SM
particles; in this case the new sector might only be probed via GWs.

The eLISA interferometer is also in a good position to probe strong
first-order phase transitions taking place well above the electroweak scale, in the
multi-TeV regime. There are many theoretical motivations for such scenarios, for 
example solving the hierarchy problem or explaining the observed dark matter abundance in our Universe. 
In fact, we have shown that eLISA may be sensitive to cosmological phase transitions taking place at
temperatures above 10 TeV, and therefore be able to probe new physics
that will remain inaccessible to collider experiments for the foreseeable  future.

We have demonstrated the extent to which different eLISA configurations can realistically detect the stochastic gravitational 
wave background arising from strong first-order cosmological phase transitions. To do so, 
we considered a (model-independent) parametrization of the phase transition in terms of $\alpha$, 
$\beta/H_*$, $T_*$, and $\alpha_{\infty}$.
To make contact with realistic models, we analysed the predicted values for these parameters 
in various well-motivated scenarios.

Focusing on the eLISA designs 
listed in Table~\ref{tab:configurations}, and taking into account the signal-to-noise ratio
thresholds required for detection (${\rm SNR_{thr}}  > 10$ for six links, ${\rm SNR_{thr}} >
50$ for four links~\cite{AntoineP}), we conclude that:
\begin{itemize}
\item For typical electroweak phase transitions, in which all relevant dimensionful parameters are near the electroweak scale, the configuration
  C1 is definitively better than the others, while the
  configuration C4 is decisively unsatisfactory. The
  performances of the configurations C2 and C3 are similar, but C2 has the ability to test a larger fraction of the considered benchmark points. 
\item For phase transitions not strictly related to electroweak
  symmetry breaking, e.g.~involving a dark sector or
  a dilaton, the predicted characteristics of the phase transitions exhibit more variation. Larger GW signals are allowed than in the electroweak case. For certain scenarios, the resulting GW
  spectrum can be probed by all considered eLISA designs. 
  The C1 configuration, however, has the potential to test a much wider region of the parameter space in such models than do C2--C4.
\end{itemize}
It is worth reiterating that our results for the four-link
configurations depend strongly on the assumed prior knowledge of the noise level,
as well as the feasibility of the data analysis technique proposed
in~\cite{Adams:2013qma}.

Finally, we emphasize that our results can be straightforwardly extended beyond
the specific models considered in Section~\ref{sec:models}. 
To do so, one should identify the main features of the bubble wall
dynamics from the general considerations of Section~\ref{sec:scenarios} and e.g.~Refs.~\cite{Bodeker:2009qy, Espinosa:2010hh, Konstandin:2010dm, Kozaczuk:2015owa},
and compute the predicted values for the 
parameters $\alpha$, $\beta/H_*$, $T_*$ and $\alpha_{\infty}$,
as defined in Sections~\ref{sec:def} and~\ref{sec:signal} (in some cases only a subset of these quantities will be relevant for the predicted GW signal).
The results of this procedure can then be compared directly with
the appropriate eLISA sensitivity curves provided in Figs.~\ref{fig:case1}--\ref{fig:var_aInf}. In this way, we
hope this study to serve as a useful tool (and motivation) for future investigations of 
eLISA's potential to probe new physics scenarios predicting
strong first-order phase transitions in the early Universe.

\subsection*{Acknowledgements}

The authors would like to thank the CERN Theory Group and the University of Stavanger for hosting the Cosmology Working Group meetings. We thank Jacopo Ghiglieri and Anders Tranberg for contributing to the discussions which led to this work.  GN was partly
supported by the Swiss National Science Foundation (SNF) under grant 200020-155935. JK is supported by the Natural Sciences and Engineering Research Council of Canada (NSERC). TK and GN were supported by the German Science Foundation (DFG)
under the Collaborative Research Center (SFB) 676 Particles, Strings and
the Early Universe. MH and SH acknowledge support from the Science and Technology Facilities Council (grant number ST/J000477/1). DJW is supported by the People Programme (Marie Sk{\l}odowska-Curie actions) of the European Union Seventh Framework Programme (FP7/2007-2013) under grant agreement number PIEF-GA-2013-629425.
We acknowledge PRACE for awarding us access to resource HAZEL HEN based in Germany at the High Performance Computing Center Stuttgart (HLRS). Our simulations also made use of facilities at the Finnish Centre for Scientific Computing CSC, and the COSMOS Consortium supercomputer (within the DiRAC Facility jointly funded by STFC and the Large Facilities Capital Fund of BIS). J.M.N. is supported by the People Programme (Marie curie Actions) of the European Union Seventh Framework Programme (FP7/2007-2013) under REA grant agreement PIEF-GA-2013-625809. 

\appendix
\section{\sc  Sensitivity to Runaway Bubbles in a Plasma}
\label{app:alphainfty}

\begin{figure}
\begin{center}
\includegraphics[width=0.32\textwidth]{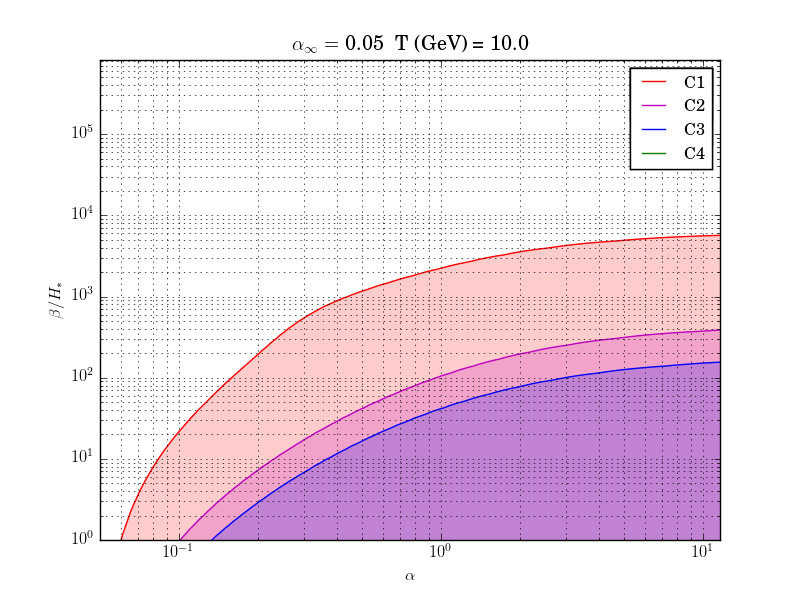}
\includegraphics[width=0.32\textwidth]{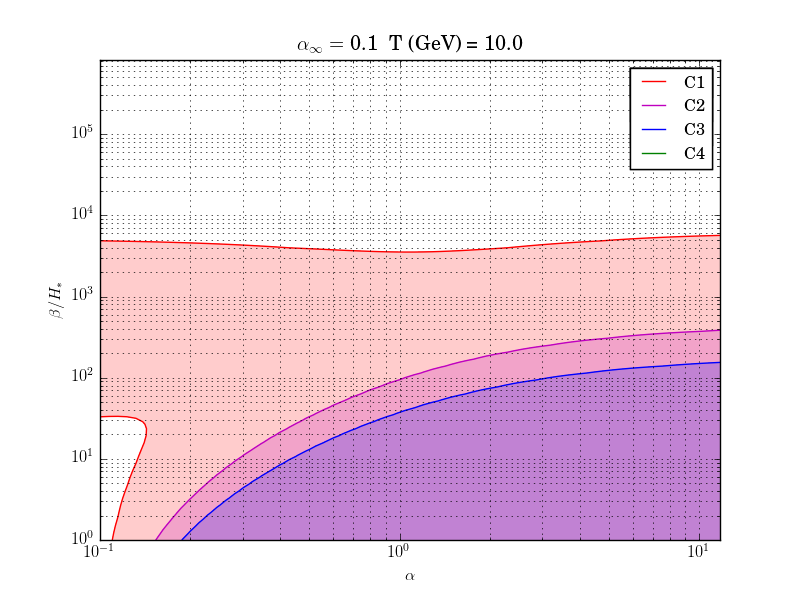}
\includegraphics[width=0.32\textwidth]{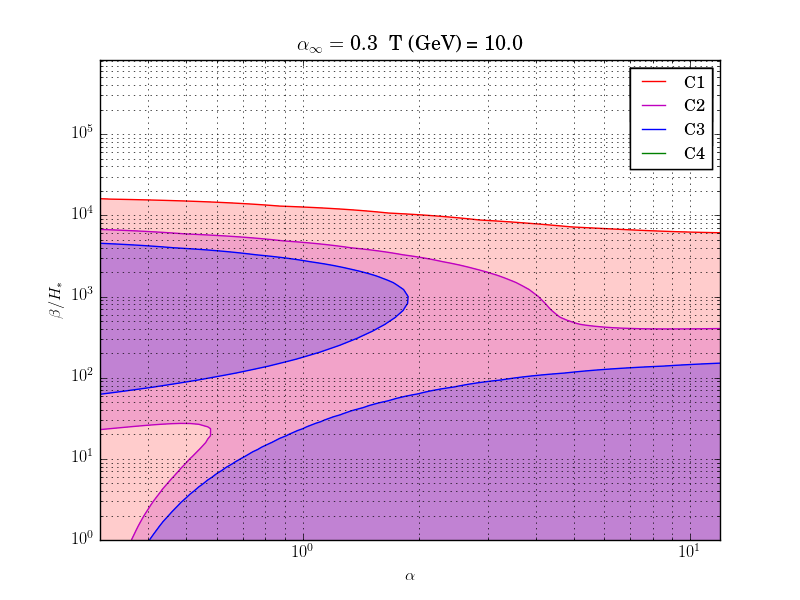}
\includegraphics[width=0.32\textwidth]{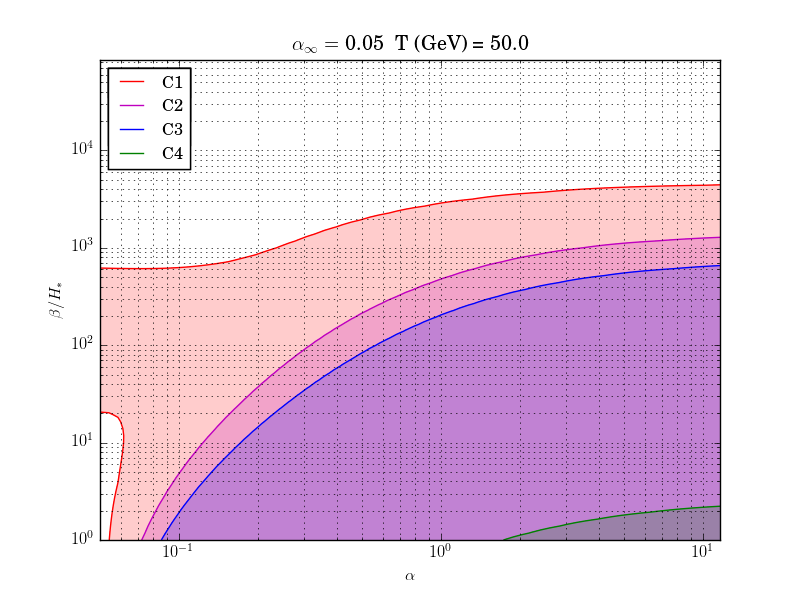}
\includegraphics[width=0.32\textwidth]{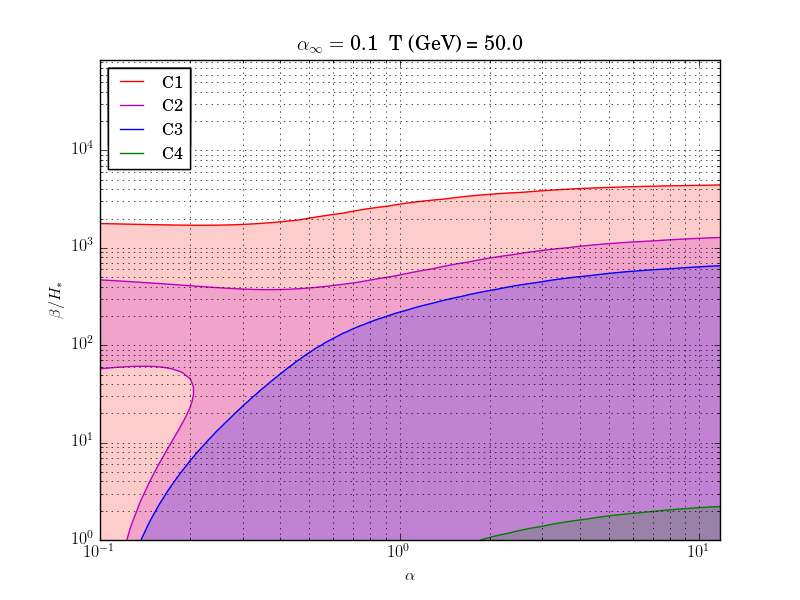}
\includegraphics[width=0.32\textwidth]{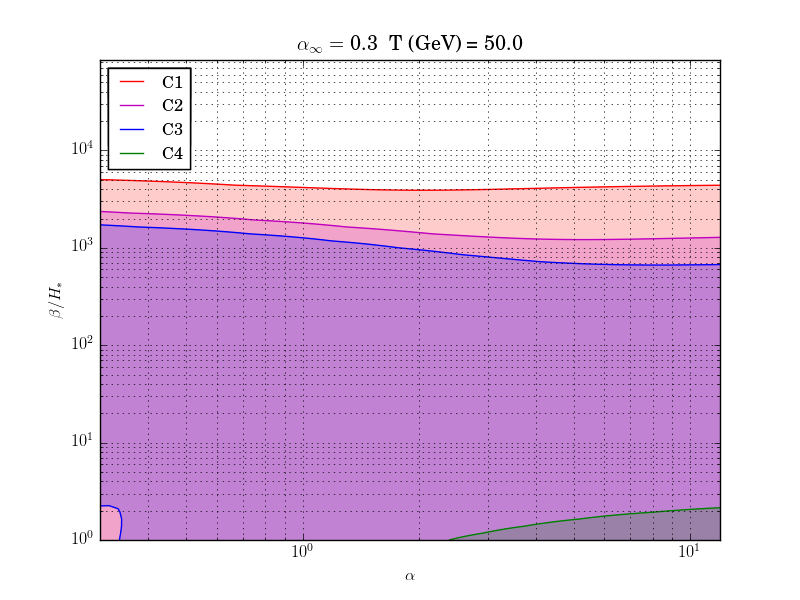}
\includegraphics[width=0.32\textwidth]{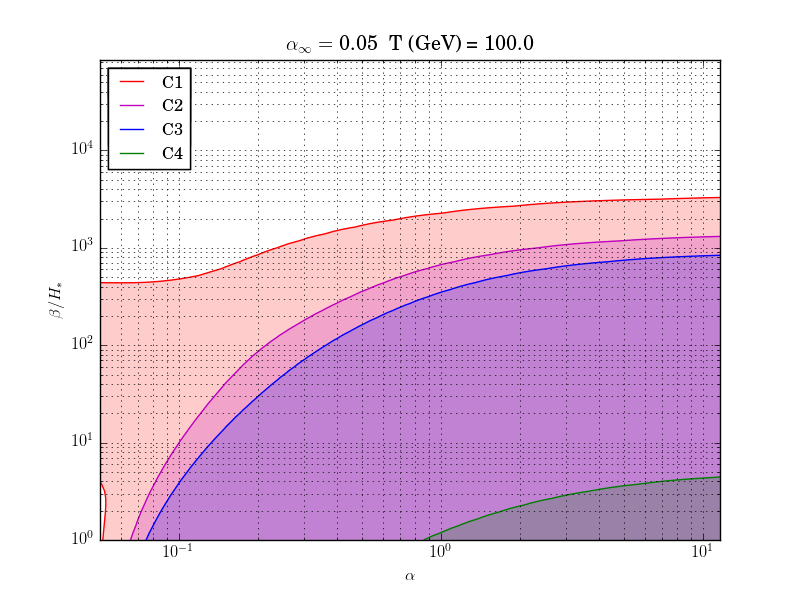}
\includegraphics[width=0.32\textwidth]{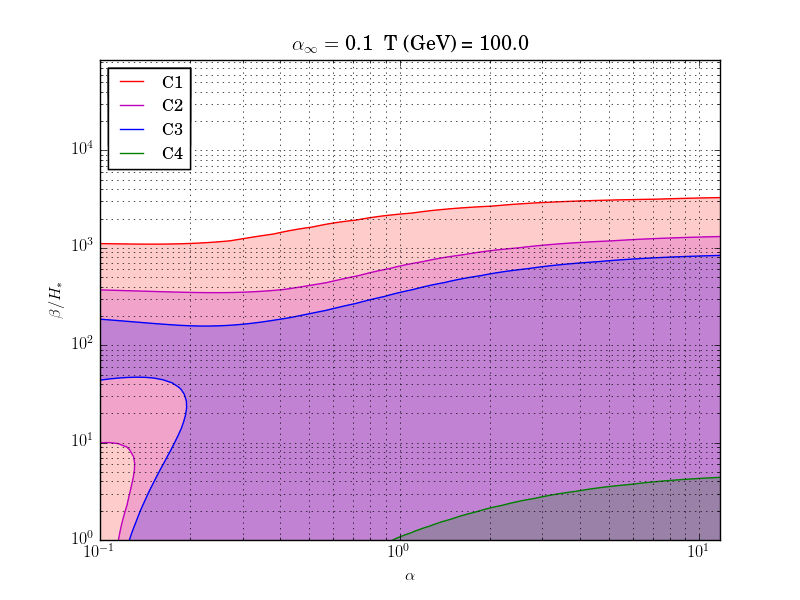}
\includegraphics[width=0.32\textwidth]{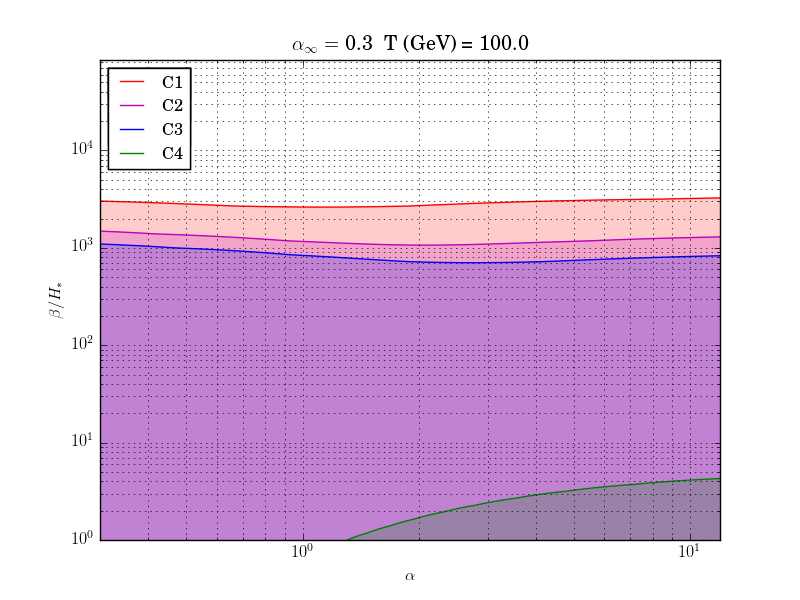}
\end{center}
\caption{\small \label{fig:var_aInf} Detectable regions in the
  ($\alpha$, $\beta/H_*$) plane for phase transitions with runaway
  bubbles at finite $\alpha$ (Case 2). In each row of panels we fix $T_*$ as indicated
  and consider three values of $\alpha_\infty$, increasing from left to right.}
\end{figure}

If the PT proceeds as in Case 2, with runaway bubbles (i.e.~$v_w=1$) at finite $\alpha$, the contours of the detectable region in the ($\alpha$, $\beta/H_*$) plane depend not only on $T_*$ and $v_w$, but also on $\alpha_\infty$. In our analysis we have consistently set $v_w$ to values very close to one, maximizing its effect on the GW spectrum. In this Appendix we aim to show how the sensitivity contours change with $T_*$ and $\alpha_\infty$. We refer the reader to Figure~\ref{fig:var_aInf} where we plot the detectable regions for three values of $T_*$ and $\alpha_\infty$. 

From Figure~\ref{fig:var_aInf} it is apparent that the effect of increasing $T_*$ is similar to that found by increasing $\alpha_\infty$. This is because, in both cases, the contribution of the sound waves to the GW spectrum gains importance. Fixing $\alpha_\infty$ amounts to fixing the relative contribution of the scalar field and of the sound waves, c.f.~Eqs.~\eqref{eq:kapv}. At small $\alpha_\infty$ and small $T_*$ only the scalar field contribution is relevant: the typical contour-shape is that appearing in the top left panel of Figure~\ref{fig:var_aInf}. For small $\beta/H_*$, only the high frequency tail $1/f$ lies within the eLISA sensitivity region. As $\beta/H_*$ increases, a larger portion of the spectrum becomes observable until the peak finally exits the detectable region. Increasing $\beta/H_*$ also causes the overall amplitude to diminish and therefore one needs correspondingly lager $\alpha$ to guarantee detectability. Starting from the situation with small $\alpha_\infty$ and small $T_*$ (top left panel), one has that increasing $T_*$ at fixed $\alpha_\infty$ does not change the relative contribution of the scalar field and sound waves but shifts the peak to higher frequencies. Since the peak due to the sound waves occurs at parametrically larger frequencies than that due to the scalar field (c.f.~Eqs.~\eqref{eq:envPeak} and~\eqref{eq:SWPeak}), this can help shift the spectrum into the detectable region, especially if $\beta/H_*$ is large enough. We reiterate that the amplitude of the sound wave contribution is larger by a factor $\beta/H_*$ than that of the scalar field in our treatment (c.f.~Eqs.~\eqref{eq:Omenv} and~\eqref{eq:OmGsound}). As a consequence, the regions at small $\alpha$ and large $\beta/H_*$ open up with increasing $T_*$, in the middle left and bottom left panels of Figure~\ref{fig:var_aInf}.

The same trend is observed at fixed $T_*$ if one increases $\alpha_\infty$: the sound wave contribution is boosted in amplitude w.r.t.~that of the scalar field and affects the detectable regions at small $\alpha$ and large $\beta/H_*$. Note that the amplitude of the scalar field contribution, Eq.~\eqref{eq:Omenv}, increases with increasing $\alpha$ until the $\alpha-$dependence drops out at very large $\alpha$ and consequently the contours demarcating the detectable regions flatten at high $\alpha$. On the other hand, the amplitude of the sound wave contribution at fixed $\alpha_\infty$ decays with growing $\alpha$ (c.f.~Eqs.~\eqref{eq:OmGsound} and~\eqref{eq:kapv}). This behaviour is reflected in the shape of the contours in the low-$\alpha$ regions shown in the right panels.


\begin{thebibliography}{99}


\bibitem{elisaweb}
https://www.elisascience.org 

\bibitem{pathfinderweb}
http://sci.esa.int/lisa-pathfinder/


\bibitem{Klein:2015hvg}
  A.~Klein {\it et al.},
  arXiv:1511.05581 [gr-qc].

\bibitem{Grojean:2006bp} 
  C.~Grojean and G.~Servant,
  Phys.\ Rev.\ D {\bf 75}, 043507 (2007)
  [hep-ph/0607107].

\bibitem{Turner:1992tz}
  M.~S.~Turner, E.~J.~Weinberg and L.~M.~Widrow,
  Phys.\ Rev.\ D {\bf 46} (1992) 2384.



\bibitem{Binetruy:2012ze}
  P.~Binetruy, A.~Bohe, C.~Caprini and J.~F.~Dufaux,
  JCAP {\bf 1206} (2012) 027
  [arXiv:1201.0983 [gr-qc]].


\bibitem{Megevand:2013yua}
  A.~Megevand and F.~A.~Membiela,
  Phys.\ Rev.\ D {\bf 89} (2014) 10,  103507
  [arXiv:1311.2453 [astro-ph.CO]].


\bibitem{Megevand:2014yua}
  A.~Megevand and F.~A.~Membiela,
  Phys.\ Rev.\ D {\bf 89} (2014) 10,  103503
  [arXiv:1402.5791 [astro-ph.CO]].

\bibitem{Espinosa:2010hh}
  J.~R.~Espinosa, T.~Konstandin, J.~M.~No and G.~Servant,
  JCAP {\bf 1006} (2010) 028
  [arXiv:1004.4187 [hep-ph]].

\bibitem{Bodeker:2009qy}
  D.~Bodeker and G.~D.~Moore,
  JCAP {\bf 0905} (2009) 009
  [arXiv:0903.4099 [hep-ph]].


\bibitem{Kosowsky:1991ua}
  A.~Kosowsky, M.~S.~Turner and R.~Watkins,
  Phys.\ Rev.\ D {\bf 45} (1992) 4514.

\bibitem{Kosowsky:1992rz}
  A.~Kosowsky, M.~S.~Turner and R.~Watkins,
  Phys.\ Rev.\ Lett.\  {\bf 69} (1992) 2026.
  
\bibitem{astro-ph/9211004}
  A.~Kosowsky and M.~S.~Turner,
  Phys.\ Rev.\ D\ {\bf 47} (1993) 4372
  [astro-ph/9211004].

\bibitem{astro-ph/9310044}
  M.~Kamionkowski, A.~Kosowsky and M.~S.~Turner,
  Phys.\ Rev.\ D\ {\bf 49} (1994) 2837
  [astro-ph/9310044].

\bibitem{Caprini:2007xq}
  C.~Caprini, R.~Durrer and G.~Servant,
  Phys.\ Rev.\ D {\bf 77} (2008) 124015
  [arXiv:0711.2593 [astro-ph]].

\bibitem{Huber:2008hg}
  S.~J.~Huber and T.~Konstandin,
  JCAP {\bf 0809} (2008) 022
  [arXiv:0806.1828 [hep-ph]].

\bibitem{Hindmarsh:2013xza}
  M.~Hindmarsh, S.~J.~Huber, K.~Rummukainen and D.~J.~Weir,
  Phys.\ Rev.\ Lett.\  {\bf 112} (2014) 041301
  [arXiv:1304.2433 [hep-ph]].

\bibitem{Giblin:2013kea}
  J.~T.~Giblin, Jr. and J.~B.~Mertens,
  JHEP {\bf 1312} (2013) 042
  [arXiv:1310.2948 [hep-th]].


\bibitem{Giblin:2014qia}
  J.~T.~Giblin and J.~B.~Mertens,
  Phys.\ Rev.\ D {\bf 90} (2014) 2,  023532
  [arXiv:1405.4005 [astro-ph.CO]].


\bibitem{Hindmarsh:2015qta}
  M.~Hindmarsh, S.~J.~Huber, K.~Rummukainen and D.~J.~Weir,
  Phys.\ Rev.\ D {\bf 92} (2015) 12,  123009
  arXiv:1504.03291 [astro-ph.CO].

\bibitem{Caprini:2006jb}
  C.~Caprini and R.~Durrer,
  Phys.\ Rev.\ D {\bf 74} (2006) 063521
  [astro-ph/0603476].
  
\bibitem{Kahniashvili:2008pf}
  T.~Kahniashvili, A.~Kosowsky, G.~Gogoberidze and Y.~Maravin,
  Phys.\ Rev.\ D {\bf 78} (2008) 043003
  [arXiv:0806.0293 [astro-ph]].

\bibitem{Kahniashvili:2008pe}
  T.~Kahniashvili, L.~Campanelli, G.~Gogoberidze, Y.~Maravin and B.~Ratra,
  Phys.\ Rev.\ D {\bf 78} (2008) 123006
   [Phys.\ Rev.\ D {\bf 79} (2009) 109901]
  [arXiv:0809.1899 [astro-ph]].

\bibitem{Kahniashvili:2009mf}
  T.~Kahniashvili, L.~Kisslinger and T.~Stevens,
  Phys.\ Rev.\ D {\bf 81} (2010) 023004
  [arXiv:0905.0643 [astro-ph.CO]].
    
\bibitem{arXiv:0909.0622}
  C.~Caprini, R.~Durrer and G.~Servant,
  JCAP\ {\bf 0912} (2009) 024
  [arXiv:0909.0622 [astro-ph.CO]].
  
\bibitem{Kisslinger:2015hua}
  L.~Kisslinger and T.~Kahniashvili,
  Phys.\ Rev.\ D {\bf 92} (2015) 4,  043006
  [arXiv:1505.03680 [astro-ph.CO]].
  
\bibitem{Child:2012qg}
  H.~L.~Child and J.~T.~Giblin, Jr.,
  JCAP {\bf 1210} (2012) 001
  [arXiv:1207.6408 [astro-ph.CO]].

\bibitem{Caprini:2009fx}
  C.~Caprini, R.~Durrer, T.~Konstandin and G.~Servant,
  Phys.\ Rev.\ D {\bf 79} (2009) 083519
  [arXiv:0901.1661 [astro-ph.CO]].


\bibitem{Pen:2015qta}
  U.~L.~Pen and N.~Turok,
  arXiv:1510.02985 [astro-ph.CO].

\bibitem{Kosowsky:2001xp}
  A.~Kosowsky, A.~Mack and T.~Kahniashvili,
  Phys.\ Rev.\ D {\bf 66} (2002) 024030
  [astro-ph/0111483].

  \bibitem{Konstandin:2010dm} 
  T.~Konstandin and J.~M.~No,
  JCAP {\bf 1102} (2011) 008
  [arXiv:1011.3735 [hep-ph]].
 
 \bibitem{Kozaczuk:2015owa} 
  J.~Kozaczuk,
  JHEP {\bf 1510} (2015) 135 
  [arXiv:1506.04741 [hep-ph]].
  
  
\bibitem{AntoineP}  
A.~Petiteau, in preparation.
 
   \bibitem{Thrane:2013oya}
  E.~Thrane and J.~D.~Romano,
  Phys.\ Rev.\ D {\bf 88} (2013) 12,  124032
  [arXiv:1310.5300 [astro-ph.IM]].
 
  \bibitem{Adams:2010vc}
  M.~R.~Adams and N.~J.~Cornish,
  Phys.\ Rev.\ D {\bf 82} (2010) 022002
  [arXiv:1002.1291 [gr-qc]].

\bibitem{Adams:2013qma}
  M.~R.~Adams and N.~J.~Cornish,
  Phys.\ Rev.\ D {\bf 89} (2014) 2,  022001
  [arXiv:1307.4116 [gr-qc]].
  

\bibitem{Delaunay:2007wb} 
  C.~Delaunay, C.~Grojean and J.~D.~Wells,
  JHEP {\bf 0804} (2008) 029
  [arXiv:0711.2511 [hep-ph]].

  \bibitem{Huber:2007vva}
  S.~J.~Huber and T.~Konstandin,
  JCAP {\bf 0805} (2008) 017
  [arXiv:0709.2091 [hep-ph]].
 
 \bibitem{Kajantie:1995kf} 
  K.~Kajantie, M.~Laine, K.~Rummukainen and M.~E.~Shaposhnikov,
  Nucl.\ Phys.\ B {\bf 466} (1996) 189 
  [hep-lat/9510020].
  
  \bibitem{Kajantie:1996mn} 
  K.~Kajantie, M.~Laine, K.~Rummukainen and M.~E.~Shaposhnikov,
  Phys.\ Rev.\ Lett.\  {\bf 77} (1996) 2887
  [hep-ph/9605288].
  
\bibitem{Ghiglieri:2015nfa}
  J.~Ghiglieri and M.~Laine,
  JCAP {\bf 1507} (2015) 07,  022
  [arXiv:1504.02569 [hep-ph]].


\bibitem{Morrissey:2012db} 
  D.~E.~Morrissey and M.~J.~Ramsey-Musolf,
  New J.\ Phys.\  {\bf 14} (2012) 125003
  [arXiv:1206.2942 [hep-ph]].
  
  \bibitem{Caprini:2011uz} 
  C.~Caprini and J.~M.~No,
  JCAP {\bf 1201} (2012) 031 
  [arXiv:1111.1726 [hep-ph]].
  
  
  \bibitem{No:2011fi}
  J.~M.~No,
  Phys.\ Rev.\ D {\bf 84} (2011) 124025
  [arXiv:1103.2159 [hep-ph]].




\bibitem{Carena} 
M.~S.~Carena, M.~Quiros and C.~E.~M.~Wagner,
  Phys.\ Lett.\  B {\bf 380} (1996) 81
  [arXiv:hep-ph/9603420].

\bibitem{Delepine} 
D.~Delepine, J.~M.~Gerard, R.~Gonzalez Felipe and J.~Weyers,
  Phys.\ Lett.\  B {\bf 386} (1996) 183
  [arXiv:hep-ph/9604440].

\bibitem{mssm08}
M.~Carena, G.~Nardini, M.~Quiros and C.~E.~M.~Wagner,
  Nucl.\ Phys.\  B {\bf 812} (2009) 243
  [arXiv:0809.3760 [hep-ph]].

\bibitem{MikkoKari}
 M.~Laine, G.~Nardini and K.~Rummukainen,
  JCAP {\bf 1301} (2013) 011
  [arXiv:1211.7344 [hep-ph]].

\bibitem{Menon:2009mz} 
  A.~Menon and D.~E.~Morrissey,
  Phys.\ Rev.\ D {\bf 79} (2009) 115020
  [arXiv:0903.3038 [hep-ph]].

\bibitem{Cohen:2012zza}
  T.~Cohen, D.~E.~Morrissey and A.~Pierce,
  Phys.\ Rev.\ D {\bf 86} (2012) 013009
  [arXiv:1203.2924 [hep-ph]].

\bibitem{Curtin:2012aa}
  D.~Curtin, P.~Jaiswal and P.~Meade,
  JHEP {\bf 1208} (2012) 005
  [arXiv:1203.2932 [hep-ph]].


\bibitem{Carena3} 
 M.~Carena, G.~Nardini, M.~Quiros and C.~E.~M.~Wagner,
  JHEP {\bf 1302} (2013) 001
  [arXiv:1207.6330 [hep-ph]];



\bibitem{Pietroni:1992in}
  M.~Pietroni,
  Nucl.\ Phys.\ B {\bf 402} (1993) 27
  [hep-ph/9207227].

\bibitem{Davies:1996qn}
  A.~T.~Davies, C.~D.~Froggatt and R.~G.~Moorhouse,
  Phys.\ Lett.\ B {\bf 372} (1996) 88
  [hep-ph/9603388].

\bibitem{Apreda:2001us}
  R.~Apreda, M.~Maggiore, A.~Nicolis and A.~Riotto,
  Nucl.\ Phys.\ B {\bf 631} (2002) 342
  [gr-qc/0107033].


\bibitem{Kozaczuk:2013fga}
  J.~Kozaczuk, S.~Profumo and C.~L.~Wainwright,
  Phys.\ Rev.\ D {\bf 87} (2013) 7,  075011
  [arXiv:1302.4781 [hep-ph]].

\bibitem{Huang:2014ifa}
  W.~Huang, Z.~Kang, J.~Shu, P.~Wu and J.~M.~Yang,
  Phys.\ Rev.\ D {\bf 91} (2015) 2,  025006
  [arXiv:1405.1152 [hep-ph]].

\bibitem{Kozaczuk:2014kva}
  J.~Kozaczuk, S.~Profumo, L.~S.~Haskins and C.~L.~Wainwright,
  JHEP {\bf 1501} (2015) 144
  [arXiv:1407.4134 [hep-ph]].


\bibitem{GWnmssm}
S.~J.~Huber, T.~Konstandin, G.~Nardini and I.~Rues,
  JCAP {\bf 1603} (2016) no.03,  036
  [arXiv:1512.06357 [hep-ph]].











  \bibitem{Craig:2014lda} 
  N.~Craig, H.~K.~Lou, M.~McCullough and A.~Thalapillil,
  arXiv:1412.0258 [hep-ph].
  
  \bibitem{Profumo:2014opa}
  S.~Profumo, M.~J.~Ramsey-Musolf, C.~L.~Wainwright and P.~Winslow,
  Phys.\ Rev.\ D {\bf 91} (2015) 3,  035018
  [arXiv:1407.5342 [hep-ph]].
  
  \bibitem{Craig:2013xia}
  N.~Craig, C.~Englert and M.~McCullough,
  Phys.\ Rev.\ Lett.\  {\bf 111} (2013) 12,  121803
  [arXiv:1305.5251 [hep-ph]].
  
\bibitem{Espinosa:2011ax} 
  J.~R.~Espinosa, T.~Konstandin and F.~Riva,
  Nucl.\ Phys.\ B {\bf 854} (2012) 592
  [arXiv:1107.5441 [hep-ph]].
  
  \bibitem{Chen:2014ask} 
  C.~Y.~Chen, S.~Dawson and I.~M.~Lewis,
  Phys.\ Rev.\ D {\bf 91} (2015) 3,  035015
  [arXiv:1410.5488 [hep-ph]].
  
  \bibitem{Curtin:2014jma} 
  D.~Curtin, P.~Meade and C.~T.~Yu,
  JHEP {\bf 1411} (2014) 127
  [arXiv:1409.0005 [hep-ph]].
  
  
  \bibitem{Fromme:2006cm}
  L.~Fromme, S.~J.~Huber and M.~Seniuch,
  JHEP {\bf 0611} (2006) 038
  [hep-ph/0605242].
  


\bibitem{Kakizaki:2015wua}
  M.~Kakizaki, S.~Kanemura and T.~Matsui,
  Phys.\ Rev.\ D {\bf 92} (2015) 11,  115007
  [arXiv:1509.08394 [hep-ph]].
  
 
  \bibitem{Dorsch:2013wja}
  G.~C.~Dorsch, S.~J.~Huber and J.~M.~No,
  JHEP {\bf 1310} (2013) 029
  [arXiv:1305.6610 [hep-ph]].
  
   
  \bibitem{Dorsch:2014qja}
  G.~C.~Dorsch, S.~J.~Huber, K.~Mimasu and J.~M.~No,
  Phys.\ Rev.\ Lett.\  {\bf 113} (2014) 21,  211802
  [arXiv:1405.5537 [hep-ph]].
  
  \bibitem{GW2HDM}
 G.~C.~Dorsch, S.~J.~Huber, K.~Mimasu and J.~M.~No,
  arXiv:1601.04545 [hep-ph].
  
  \bibitem{Patel:2013zla} 
  H.~H.~Patel, M.~J.~Ramsey-Musolf and M.~B.~Wise,
  Phys.\ Rev.\ D {\bf 88} (2013) 1, 015003 
  [arXiv:1303.1140 [hep-ph]].
  


\bibitem{Grojean:2004xa} 
  C.~Grojean, G.~Servant and J.~D.~Wells,
  Phys.\ Rev.\ D {\bf 71} (2005) 036001 
  [hep-ph/0407019].
  

\bibitem{Nardini:2007me}
  G.~Nardini, M.~Quiros and A.~Wulzer,
  JHEP {\bf 0709} (2007) 077
  [arXiv:0706.3388 [hep-ph]].


\bibitem{Konstandin:2011dr}
  T.~Konstandin and G.~Servant,
  JCAP {\bf 1112} (2011) 009
  [arXiv:1104.4791 [hep-ph]].

  
\bibitem{Randall:1999ee} 
  L.~Randall and R.~Sundrum,
  Phys.\ Rev.\ Lett.\  {\bf 83} (1999) 3370
  [hep-ph/9905221].

\bibitem{Goldberger:1999uk} 
  W.~D.~Goldberger and M.~B.~Wise,
  Phys.\ Rev.\ Lett.\  {\bf 83} (1999) 4922 
  [hep-ph/9907447].

\bibitem{Rattazzi:2000hs}
  R.~Rattazzi and A.~Zaffaroni,
  JHEP {\bf 0104} (2001) 021
  [hep-th/0012248].

\bibitem{Garriga:2002vf} 
  J.~Garriga and A.~Pomarol,
  Phys.\ Lett.\ B {\bf 560} (2003) 91
  [hep-th/0212227].

\bibitem{Creminelli:2001th}
  P.~Creminelli, A.~Nicolis and R.~Rattazzi,
  JHEP {\bf 0203} (2002) 051
  [arXiv:hep-th/0107141].

\bibitem{Randall:2006py}
  L.~Randall and G.~Servant,
  JHEP {\bf 0705} (2007) 054
  [arXiv:hep-ph/0607158].

\bibitem{Konstandin:2010cd}
  T.~Konstandin, G.~Nardini and M.~Quiros,
  Phys.\ Rev.\ D {\bf 82} (2010) 083513
  [arXiv:1007.1468 [hep-ph]].


\bibitem{Servant:2014bla} 
  G.~Servant,
  Phys.\ Rev.\ Lett.\  {\bf 113} (2014) 17, 171803
  [arXiv:1407.0030 [hep-ph]].
  
\bibitem{Bai:2013xga}
  Y.~Bai and P.~Schwaller,
  Phys.\ Rev.\ D {\bf 89} (2014) 6,  063522
  [arXiv:1306.4676 [hep-ph]].
  
\bibitem{Boddy:2014yra}
  K.~K.~Boddy, J.~L.~Feng, M.~Kaplinghat and T.~M.~P.~Tait,
  Phys.\ Rev.\ D {\bf 89} (2014) 11,  115017
  [arXiv:1402.3629 [hep-ph]].
  
\bibitem{Schwaller:2015tja}
  P.~Schwaller,
  Phys.\ Rev.\ Lett.\  {\bf 115} (2015) 18,  181101
  [arXiv:1504.07263 [hep-ph]].

\end{thebibliography}
\end{document}